\documentclass[sigconf,nonacm]{acmart} 
\usepackage{booktabs}
\usepackage{tabularx}
\usepackage{amsmath}
\usepackage{graphicx}
\usepackage{makecell}
\usepackage{multirow}
\usepackage{colortbl}
\usepackage{placeins}
\usepackage{float}
\usepackage{enumitem}

\usepackage{adjustbox}
\usepackage{threeparttable}

\usepackage{pifont}

\usepackage{CJKutf8}

\definecolor{checkgreen}{HTML}{18845B}
\definecolor{crossred}{HTML}{C4473A}

\newcommand{\cmark}{\textcolor{checkgreen}{\ding{51}}}
\newcommand{\xmark}{\textcolor{crossred}{\ding{55}}}
\newcommand{\dataset}{\textsc{NoteHabitat}}

\definecolor{reasonblue}{HTML}{2B6CB0}
\definecolor{oursbg}{HTML}{EAF5EF}
\definecolor{resultshade1}{HTML}{FCFBFF}
\definecolor{resultshade2}{HTML}{F3EFFC}
\definecolor{resultshade3}{HTML}{E7DFF9}
\definecolor{resultshade4}{HTML}{D7CAF6}
\definecolor{resultshade5}{HTML}{C6B5F2}
\newcommand{\bestval}[1]{\textbf{#1}}
\newcommand{\secondval}[1]{\underline{#1}}
\setcopyright{acmlicensed}
\copyrightyear{2027}
\acmConference[WWW '27]{The ACM Web Conference 2027}{May 10--14, 2027}{Dublin, Ireland}
\acmYear{2027}
\acmDOI{}
\acmISBN{}

\begin{document}
\raggedbottom
\title[No Note Is an Island: Probing LLMs' Understanding of Crowdsourced Fact-Checking Environments]{No Note Is an Island: Probing LLMs' Understanding \\ of Crowdsourced Fact-Checking Environments}
\author{Weijian Fan}
\authornote{These authors contributed equally to this work.}
\email{wjfan@cuc.edu.cn}
\affiliation{
  \institution{Communication University of China}
  \city{Beijing}
  \country{China}
}

\author{Chichen Lin}
\authornotemark[1]
\email{cuc\_lcc@cuc.edu.cn}
\affiliation{
  \institution{Communication University of China}
  \city{Beijing}
  \country{China}
}

\author{Xiao Han}
\email{hanxiao@cuc.edu.cn}
\affiliation{
  \institution{Communication University of China}
  \city{Beijing}
  \country{China}
}

\author{Jiaying Wu}
\authornote{Corresponding author.}
\email{jiayingwu@u.nus.edu}
\affiliation{
  \institution{National University of Singapore}
  \city{Singapore}
  \country{Singapore}
}

\renewcommand{\shortauthors}{Fan et al.}

\begin{abstract}
Large language models (LLMs) increasingly assist crowdsourced fact-checking, yet evaluations focus on individual annotations. A correction's reach and reception depend on visibility, audience responses, and platform policies. We introduce \textbf{correction context understanding}: assessing corrective content, interpreting audience responses, and reasoning about platform dynamics within a shared environment. We build \dataset{} from 36 months of Weibo Annotations data, linking 102,994 posts (42,853 annotations), 431,407 comments, and five policy events. Evaluating diverse model families across seven tasks spanning content, interaction, and management reveals three gaps. Annotation quality correlates only weakly with engagement ($\rho=0.068$), while four zero-shot LLMs systematically underrate quality relative to human judgments, especially evidence support and objectivity. Models also miss toxic comments supporting corrections (36.4\% average recall across ten models). At the platform level, LLMs partially predict policy-associated outcome directions, but LLM-guided simulations poorly reproduce daily activity. Cross-group analyses expose blind spots tied to contribution types and reader stances that aggregate scores conceal. Reliable fact-checking AI must understand the note, the conversation around it, and the platform in which both unfold.

\end{abstract}
\maketitle

\begin{figure}[t]
\centering\includegraphics[width=\columnwidth]{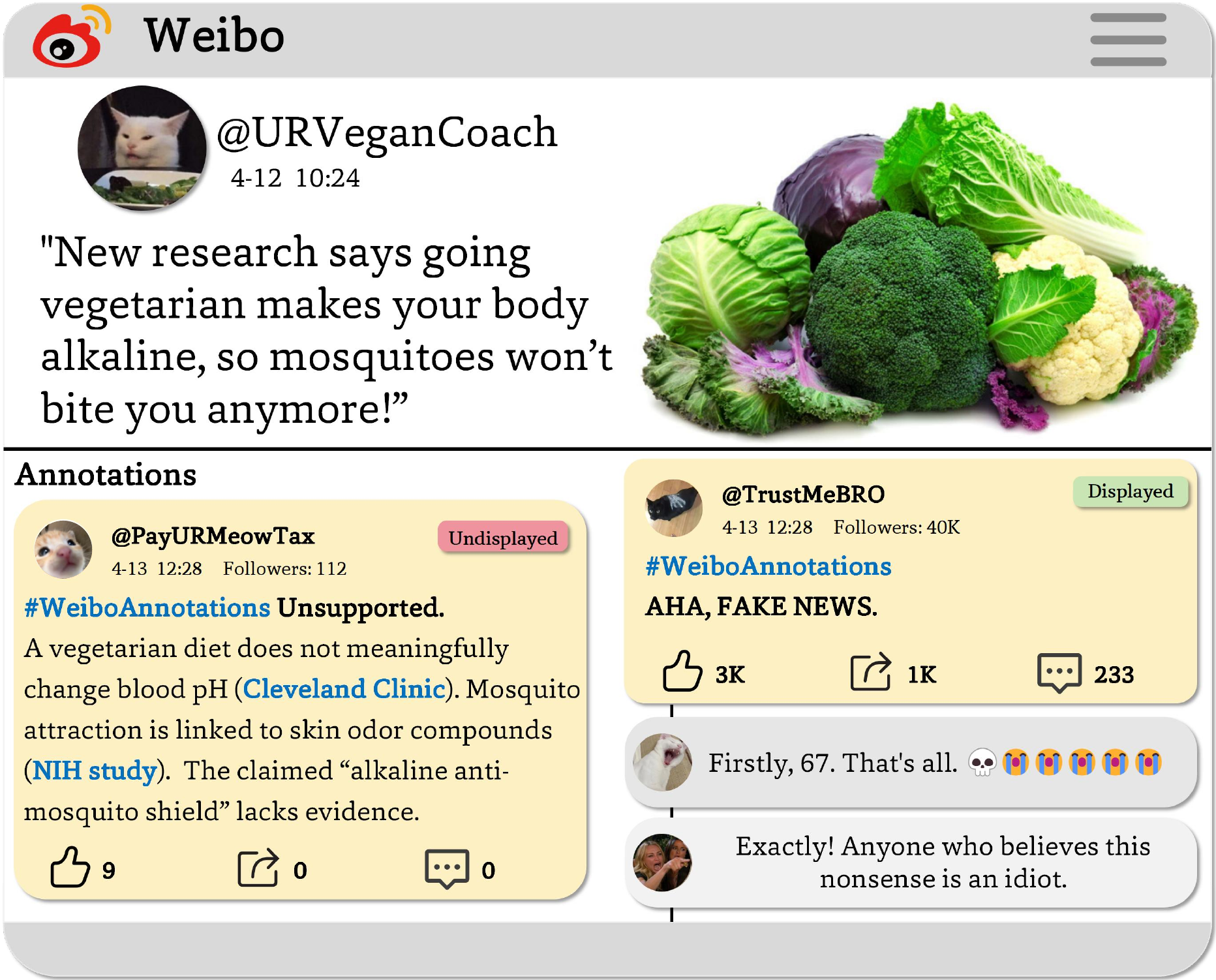}
\caption{Annotation quality, visibility, and audience interactions can diverge. Two annotations address the same misleading claim. An evidence-based correction remains undisplayed and receives little engagement, whereas an unsupported dismissal is displayed and attracts substantial attention, including an irrelevant reply and a toxic comment defending the original claim.}
\Description{An illustrated Weibo post claims that a vegetarian diet makes the body alkaline and prevents mosquito bites. The left annotation, by PayURMeowTax with 112 followers, labels the claim unsupported and cites Cleveland Clinic and an NIH study. It is undisplayed and has 9 likes, no reposts, and no comments. The right annotation, by TrustMeBRO with 40,000 followers, says AHA, FAKE NEWS. It is displayed and has 3,000 likes, 1,000 reposts, and 233 comments. Two comments appear below it: an irrelevant reply containing 67 and emojis, and an insulting reply supporting the original claim through personal experience and challenging the annotation.}
\label{fig:environment-teaser}
\end{figure}


\section{Introduction}

A well-supported fact-checking note may remain unseen, while an unsupported dismissal attracts substantial attention and hostile responses. Figure~\ref{fig:environment-teaser} illustrates this divergence: a correction's quality depends on its content, while its visibility and reception are shaped by audience behavior and platform mechanisms. As crowdsourced fact-checking systems such as X's Community Notes~\cite{wojcik2022birdwatch} and Weibo Annotations~\cite{weibolaunch} incorporate LLM-assisted writing tools~\cite{communitywriter2026}, understanding this distinction becomes essential for reliable AI assistance. Accurate content assessment alone cannot determine whether a correction needs revision, greater visibility, or protection from harmful interactions.

The theory of observed correction~\cite{bodevraga2025theory} emphasizes visibility and association with the original claim as conditions for encountering corrections. Empirical studies further show that delayed display can limit reductions in misinformation sharing~\cite{chuai2026spread}, surrounding comments can weaken corrective effectiveness~\cite{e2026comments}, and hostile feedback may discourage participation~\cite{moser2026hostile}. Nevertheless, existing fact-checking benchmarks primarily assess annotation helpfulness and quality~\cite{communitynotes,comrate,supernotes,healthnotes}. Related AI research examines audience interactions and platform governance independently~\cite{huang2025legitimacy,cho2024moderators,huang2026policysim}. These studies illuminate individual components but leave their relationships largely unexamined in AI evaluation. This gap motivates a broader question: \textbf{How well do LLMs reason about corrective contributions and the social and platform conditions that shape their reception?}

We define this capability as \textbf{correction context understanding}: the ability to assess corrective content, interpret audience responses, and reason about platform dynamics within a shared fact-checking environment. To investigate it, we introduce \dataset{}, a benchmark built on Weibo Annotations, where user-submitted corrections undergo platform review and distribution. We construct a 36-month archive comprising 102,994 annotation-related posts, including 42,853 annotations and 431,407 comments, alongside five documented policy events. The archive links corrections, source posts, reader responses, and platform history, enabling systematic evaluation of distinct capabilities and analysis of context-specific model failures within a common empirical setting.

\begin{figure*}[t]
\centering\includegraphics[width=0.95\textwidth]{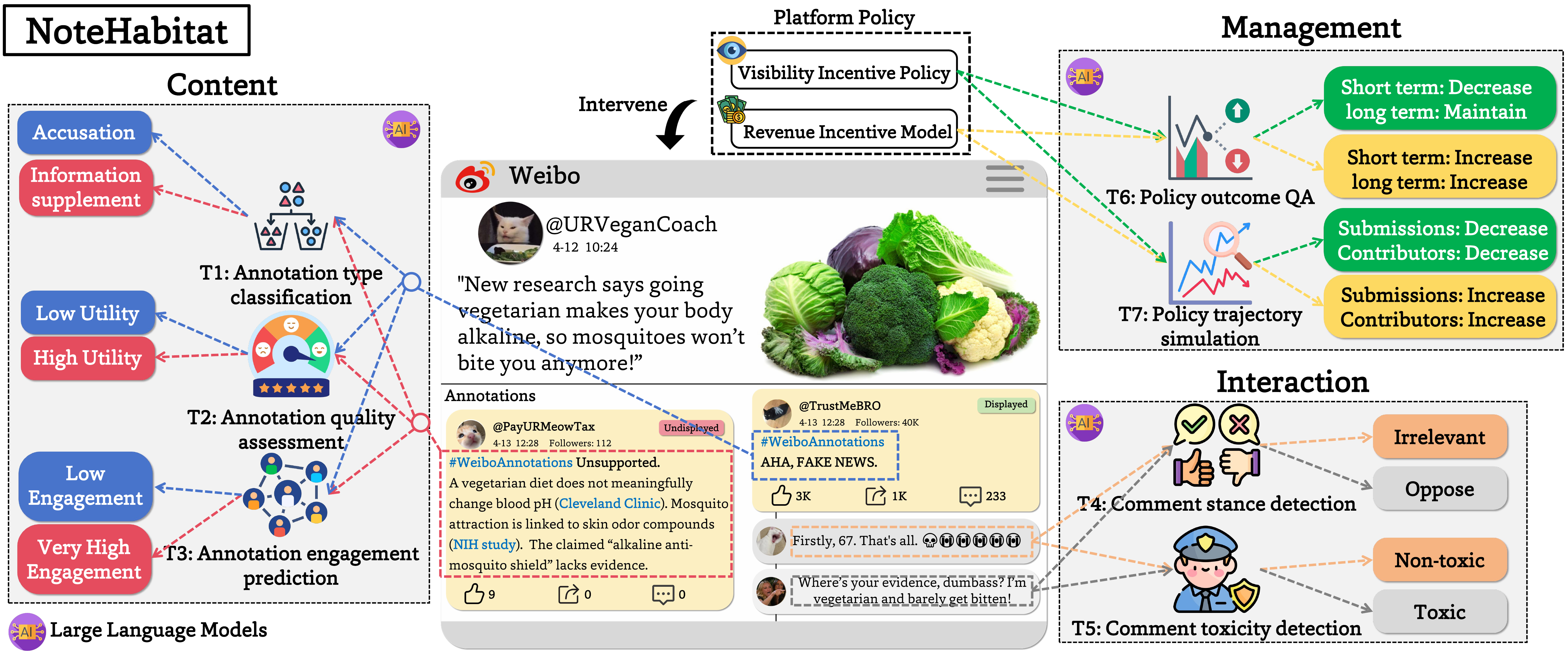}
\caption{\textbf{Seven tasks within a shared fact-checking environment.} Two annotations of the same Weibo post illustrate how \dataset{} evaluates contribution type, quality, and engagement (\textbf{Content}, T1--T3); reader stance and toxicity (\textbf{Interaction}, T4--T5); and policy-associated outcomes and activity trajectories (\textbf{Management}, T6--T7). Policy scenarios and predicted outcomes are illustrative.}
\Description{The central Weibo example shows an evidence-based information supplement and a brief accusation responding to the same claim. The content dimension includes annotation type classification, quality assessment, and engagement prediction. The interaction dimension distinguishes an irrelevant, non-toxic comment containing 67 from a toxic comment opposing the brief annotation. The management dimension uses illustrative visibility and revenue incentives to show short- and long-term policy outcome QA and simulations of submission and contributor trajectories.}
\label{fig:process-tasks}
\end{figure*}

\dataset{} organizes seven tasks into three dimensions (Figure~\ref{fig:process-tasks}). \textbf{Content} evaluates annotation type classification (T1), quality assessment (T2), and engagement prediction (T3). \textbf{Interaction} examines comment stance (T4) and toxicity (T5). \textbf{Management} evaluates policy-associated outcomes (T6) and activity trajectories (T7). These tasks probe three consequential distinctions: annotation quality and audience attention, expressed support and interaction safety, and policy outcome directions and temporal dynamics. Each distinction tests whether performance on one aspect of correction reliably extends to another.

Our evaluation of classical machine learning methods, pretrained language models, and LLMs reveals limitations across all three dimensions. Annotation quality correlates only weakly with engagement ($\rho=0.068$), while four zero-shot LLMs systematically underrate quality relative to human references, particularly for evidence support and objectivity. Models frequently overlook toxic comments supporting corrections, achieving only 36.4\% average recall across ten models. At the platform level, LLMs show some ability to predict policy-associated outcome directions, but LLM-guided agent-based simulations exhibit weak temporal agreement with observed activity. Across these settings, seemingly useful predictions leave consequential aspects of correction poorly understood.

Our contributions are threefold:
\begin{itemize}[leftmargin=*]
    \item We define correction context understanding and operationalize it through seven tasks covering corrective content, audience interactions, and platform dynamics.
    \item We introduce \dataset{}, a benchmark linking 36 months of corrective contributions, reader responses, and platform policy changes within a shared evaluation environment.
    \item We evaluate diverse model families, identifying biases in annotation quality assessment, overlooked toxicity in supportive comments, and limited temporal fidelity in platform simulations.
\end{itemize}

\begin{table}[t]
\caption{\textbf{Comparison of crowdsourced fact-checking benchmarks.} \dataset{} combines fine-grained quality ratings with evaluation tasks spanning content, interaction, and management.}
\label{tab:dataset-comparison}
\centering\small
\setlength{\tabcolsep}{1.5pt}\renewcommand{\arraystretch}{1.18}
\renewcommand{\tabularxcolumn}[1]{m{#1}}
\begin{tabularx}{\columnwidth}{@{}>{\raggedright\arraybackslash}X>{\raggedright\arraybackslash}m{34pt}>{\raggedright\arraybackslash}m{40pt}>{\centering\arraybackslash}m{28pt}>{\centering\arraybackslash}m{54pt}@{}}
\toprule
\textbf{Resource} & \textbf{Platform} & \textbf{Language} & \makecell[c]{\textbf{Quality}\\\textbf{ratings}} & \makecell[c]{\textbf{Context}\\\textbf{understanding}} \\
\midrule
HelloFresh \cite{hellofresh} & X / Wikipedia & English & Coarse & \xmark \\
COMMUNITYNOTES \cite{communitynotes} & X & Multilingual & Coarse & \xmark \\
HealthNotes \cite{healthnotes} & X & English & Coarse & \xmark \\
PolBench \cite{polbench} & X & English & Coarse & \xmark \\
ComRate \cite{comrate} & X & Multilingual & Coarse & \xmark \\
Supernotes \cite{supernotes} & X & English & Fine & \xmark \\
Mohammadi et al. \cite{fouryearcn} & X & Multilingual & Coarse & \xmark \\
\midrule
\rowcolor{oursbg}
\textbf{\dataset{}} & \textbf{Weibo} & \textbf{Chinese} & \textbf{Fine} & \cmark \\
\bottomrule
\end{tabularx}
\par\smallskip\raggedright\footnotesize
Fine: separate ordinal human ratings across quality dimensions; Coarse: overall ratings. Reason tags and model scores are excluded.
\cmark: evaluated tasks cover all three context dimensions; \xmark: not all three. Raw fields alone do not count.
\end{table}

\section{Background and Related Work}
\label{sec:related}
\textbf{AI Assistance and Annotation-Centered Evaluation.}
AI assistance in crowdsourced fact-checking has largely centered on the annotation as the unit of generation and evaluation. Models support annotation writing, helpfulness prediction, and quality assessment~\cite{supernotes,healthnotes,li2025scaling,zhang2026canote,de2026feedback}, with Community Writer illustrating their growing deployment in platform practice~\cite{communitywriter2026}. Existing benchmarks similarly emphasize annotation content and rating-based judgments~\cite{communitynotes,comrate,hellofresh,polbench}. These resources establish how well models can produce or assess corrective contributions, but offer limited evaluation of their public reception and surrounding platform conditions (Table~\ref{tab:dataset-comparison}).

\vspace{0.2em}
\textbf{Social and Platform Conditions of Correction.}
In practice, corrective communication unfolds within a social and platform-mediated environment. The theory of observed correction emphasizes visibility and association with the original claim as conditions for encountering corrections~\cite{bodevraga2025theory}. Platforms shape these conditions differently. X's Community Notes relies on contributor helpfulness ratings~\cite{communitynotes}, whereas Weibo Annotations centrally reviews submissions and distributes approved annotations through multiple channels~\cite{weibofaq,weibop1}. Empirical studies show that display delays and uneven visibility can constrain corrective reach~\cite{chuai2026spread,bouchaud2026polarizing}, while surrounding comments can weaken corrective effectiveness and hostile feedback may discourage participation~\cite{e2026comments,moser2026hostile}. Subsequent user behavior and continued contribution also vary with correction history and publication opportunities~\cite{chuai2026downstream,arjmandi2025sustainability}. Research on Weibo Annotations further identifies participation, review, reach, and protection as key governance bottlenecks~\cite{weiboChiStudy}. These findings establish the importance of understanding the conditions surrounding corrective contributions.

\vspace{0.2em}

\textbf{Contextual AI Capabilities and Evaluation Gaps.}
Related AI research provides methods for examining parts of this environment. Engagement prediction models audience attention~\cite{volkovs2020engagement,xu2025smtpd}, while stance and harmful-language detection evaluate the interpretation of user responses~\cite{rumoureval,rottger2021hatecheck}. Policy outcome prediction and social simulation examine behavior under changing conditions~\cite{ashokkumar2026experiments,hut2026simulation,huang2026policysim}. However, these capabilities are generally evaluated in separate settings, with limited connection to the corrective contributions, audience responses, and platform policies of a shared fact-checking environment. To address this gap, \dataset{} links source posts, annotations, reader comments, and longitudinal policy records. Its seven tasks operationalize \textit{correction context understanding} across content, interaction, and management, exposing capability gaps that annotation-centered evaluations cannot capture.


\section{\dataset{}: Benchmarking Correction Context Understanding}
\label{sec:benchmark}\label{sec:corpus}

A correction unfolds within a platform environment. A contribution must first be identified and assessed, may or may not attract audience attention, and can generate responses shaped by platform rules. To capture these connected aspects, \dataset{} organizes seven tasks (T1--T7) into three dimensions (Figure~\ref{fig:process-tasks}). \textbf{Content} (\S\ref{sec:content}) evaluates user contribution types, their quality, and observed engagement. \textbf{Interaction} (\S\ref{sec:interaction})  examines how readers respond and whether those responses contain harmful language. \textbf{Management} (\S\ref{sec:policy-events})  assesses policy-associated outcomes and platform activity over time. The design preserves three important distinctions: contribution type, quality and audience attention, expressed stance and interaction safety, and predicted outcome directions and temporal dynamics.

\subsection{Building a Linked Fact-Checking Archive}
\label{sec:data-profile}
\begin{figure*}[t]
\centering\includegraphics[width=\textwidth]{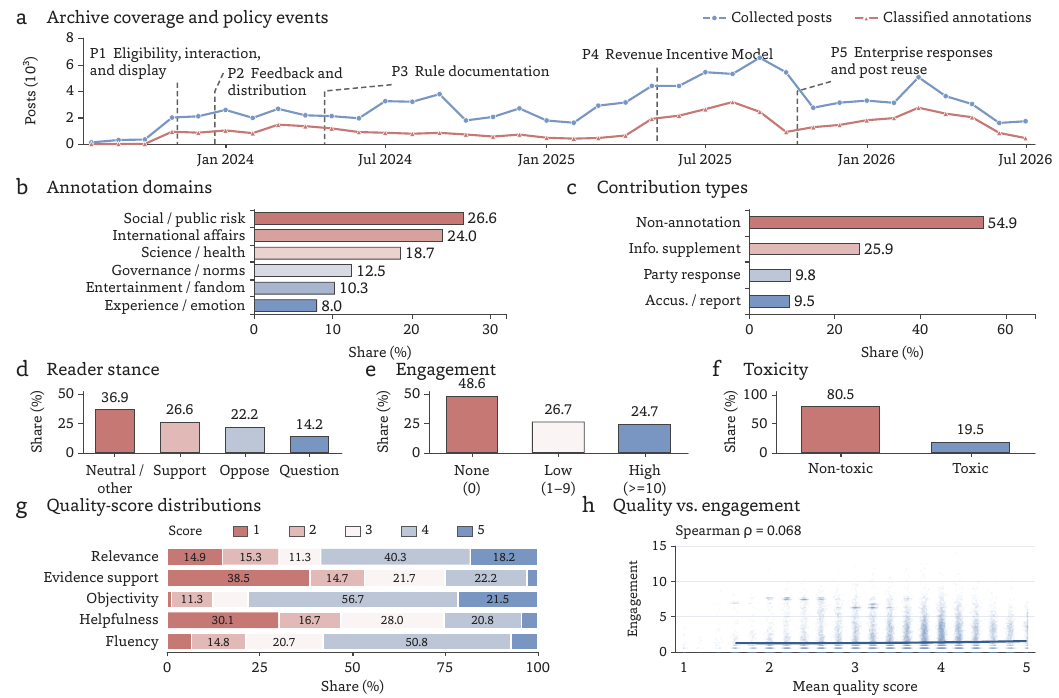}
\caption{\textbf{Platform history and benchmark composition.} (a) Monthly posts and classified annotations, with policy events P1--P5; boundary months are partial. (b--g) Distributions of annotation domains, contribution types (T1), reader stance (T4), engagement (T3), toxicity (T5), and five quality dimensions (T2), respectively. (h) Correlations between annotation quality and engagement. Each task has 24,000 instances; T4/T5 share comments. Quality distributions combine 22,000 model-labeled training/validation and 2,000 test instances.}
\Description{Seven panels show monthly archive coverage, annotation domains, contribution types, reader stance, engagement levels, toxicity labels, and five discrete quality-score frequency plots. Panel a places policy numbers and full names at staggered heights inside the plot, connected to event-date markers by uniform dashed lines: P1 Eligibility, interaction, and display on November 7, 2023; P2 Feedback and distribution on December 20, 2023; P3 Rule documentation on April 23, 2024; P4 Revenue Incentive Model on May 7, 2025; and P5 Enterprise responses and post reuse on October 14, 2025. Panel g retains all twenty-five score percentages with a common scale.}
\label{fig:dataset-profile}

\end{figure*}

\begin{CJK*}{UTF8}{gbsn}
To evaluate these capabilities in a shared environment, we collected public Weibo posts associated with the Weibo Annotations hashtag (\#微博附注) and related markers, along with metadata, linked source posts, and comments. The archive spans August 17, 2023 to July 14, 2026, covering 102,994 annotation-related posts, including 42,853 annotations, and 431,407 analyzable comments. These linked records capture corrective contributions and their reception, while the longitudinal archive documents activity around changes to platform rules and incentives. Figure~\ref{fig:dataset-profile} summarizes archive coverage and benchmark distributions.
\end{CJK*}

For T1--T5, we construct task-specific samples of 24,000 instances, each divided into 20,000 training, 2,000 validation, and 2,000 test instances. T4 and T5 share the same comments, allowing stance and toxicity to be evaluated separately on identical interactions. The annotation set was labeled using the aligned Qwen3-4B model and manually checked, with six annotators providing human references. Appendix~\ref{sec:labeling-workflow} details the labeling protocol.

\subsection{Content: Contribution Quality and Audience Attention}
\label{sec:content}

Content evaluation begins with three distinct questions. \textit{What type of contribution has been submitted? How useful is its content? How much attention does it receive?} The distinctions matter because platform records can contain material that is not itself an annotation, while a well-supported correction may attract little engagement. T1--T3 assess these properties separately, enabling analysis of where contribution quality and audience attention diverge.

\textbf{T1: Annotation type classification.}
Before assessing annotation quality, models must distinguish substantive contributions from other annotation-related posts. T1 classifies instances into \emph{information supplements}, \emph{involved-party responses} (including institutional clarifications), \emph{accusations/reports}, and \emph{non-annotations}. Non-annotations constitute 54.9\% of the task sample, making this distinction important for screening and organizing contributions. The remaining categories also capture different communicative purposes that may require different considerations during review.

\vspace{0.2em}

\textbf{T2: Annotation quality assessment.}
Annotation quality is not directly captured by review outcomes or engagement counts~\cite{weibofaq,xnotesdata}. Following prior work~\cite{supernotes,healthnotes,wu2026beyond}, T2 scores five dimensions on a 1--5 scale: \emph{relevance} (alignment with the source claim), \emph{evidence support} (supporting evidence or reasoning), \emph{objectivity} (factual consistency and qualified claims), \emph{helpfulness} (value for reader understanding), and \emph{fluency} (clarity and coherence). Dimensions are assessed independently~\cite{liu2023geval} and aggregated into an overall utility score, supporting review and targeted contributor revisions~\cite{truong2025community,weiboChiStudy}. We report MAE, RMSE, and Pearson correlation to assess rating accuracy and agreement with human judgments. Rating anchors appear in Appendix Table~\ref{tab:quality-codebook}.

\vspace{0.2em}
\textbf{T3: Annotation engagement prediction.}
T3 predicts engagement from annotation and available source-post text, defining $e$ as the sum of archived likes, comments, and reposts~\cite{vairetti2026media,srinivasan2013engaging}. Regression targets $\ln(1+e)$, while classification distinguishes \emph{none} (0), \emph{low} (1--9), and \emph{high} ($\geq10$). ML and encoder LM baselines train separate models; LLMs output regression values and class scores, selecting the highest-scoring class. Excluding 1,303 records with unavailable source links leaves 41,550 annotations eligible for temporal sampling. Figure~\ref{fig:process-tasks} contrasts an evidence-supported annotation receiving nine likes with an unsupported accusation attracting thousands of interactions, motivating separate evaluation of quality (T2) and engagement (T3) to identify valuable but overlooked contributions~\cite{weiboChiStudy}.

\subsection{Interaction: Reader Stance and Safety}
\label{sec:interaction}

Engagement counts describe the volume of audience response, but reveal little about what readers are saying. Responses may express agreement, request clarification, challenge an annotation, or introduce hostility. Importantly, agreement and safety are distinct properties: a comment supporting a correction may still contain harassment or abuse. T4 and T5 therefore evaluate stance and toxicity independently, using the same reader comments and surrounding conversational context.
\textbf{T4: Comment stance detection.}
T4 classifies a comment's position relative to the annotation as \emph{support}, \emph{question}, \emph{oppose}, or \emph{neutral/other}, following established rumour-response classification~\cite{rumoureval}. Inputs include the comment and available parent-comment, annotation, and source-post text. Distinguishing agreement from questions and objections helps characterize audience interpretation and can identify concerns that contributors may need to address.

\vspace{0.2em}
\textbf{T5: Comment toxicity detection.}
Stance alone cannot establish whether an interaction is safe. T5 identifies harmful expression by the comment author, regardless of whether the comment supports or opposes the annotation, using the same contextual inputs as T4. Hostile exchanges can expose contributors and annotated users to harassment, increasing the social costs of public correction~\cite{weiboChiStudy}. This task evaluates a complementary capability relevant to moderation and user protection, including the recognition of abusive supportive feedback.

Figure~\ref{fig:process-tasks} illustrates these distinct judgments. The ``67'' response is irrelevant and non-toxic, whereas the hostile comment challenging the unsupported accusation is classified as opposing and toxic. Evaluating both dimensions prevents favorable or unfavorable stance from being treated as a proxy for interaction safety.


\begin{table*}[!t]
\caption{\textbf{Model performance on content and interaction tasks (T1--T5).}
Each task has 2,000 test instances, with 1,822 notes shared across T1--T3.
T2 averages metrics across five quality dimensions; T3 regression uses $\ln(1+\text{engagement})$.
F1 denotes Macro-F1 except Toxic-F1 for T5. Results average three runs.
Five purple shades indicate within-column ranks (darker is better); bold and underlined values mark best and second-best, respectively, including ties.}
\label{tab:main-results}
\centering\normalfont\small
\setlength{\tabcolsep}{3pt}\renewcommand{\arraystretch}{1.08}
\begin{tabular*}{\textwidth}{@{\extracolsep{\fill}}lrrrrrrrrrrrrrr@{}}
\toprule
& \multicolumn{2}{c}{\textsc{T1: Type}} & \multicolumn{3}{c}{\textsc{T2: Quality}} & \multicolumn{5}{c}{\textsc{T3: Engagement}} & \multicolumn{2}{c}{\textsc{T4: Stance}} & \multicolumn{2}{c}{\textsc{T5: Toxicity}} \\
\cmidrule(lr){2-3}\cmidrule(lr){4-6}\cmidrule(lr){7-11}\cmidrule(lr){12-13}\cmidrule(l){14-15}
Model & Acc.$\uparrow$ & F1$\uparrow$ & MAE$\downarrow$ & RMSE$\downarrow$ & $r\uparrow$ & Acc.$\uparrow$ & F1$\uparrow$ & MAE$\downarrow$ & RMSE$\downarrow$ & $r\uparrow$ & Acc.$\uparrow$ & F1$\uparrow$ & Acc.$\uparrow$ & F1$\uparrow$ \\
\midrule
XGBoost & \cellcolor{resultshade4}.729 & \cellcolor{resultshade2}.487 & \cellcolor{resultshade3}.959 & \cellcolor{resultshade3}1.125 & \cellcolor{resultshade1}.186 & \cellcolor{resultshade3}.623 & \cellcolor{resultshade3}.588 & \cellcolor{resultshade3}1.204 & \cellcolor{resultshade4}1.676 & \cellcolor{resultshade3}.757 & \cellcolor{resultshade1}.424 & \cellcolor{resultshade1}.279 & \cellcolor{resultshade1}.792 & \cellcolor{resultshade1}.011 \\
Random Forest & \cellcolor{resultshade3}.720 & \cellcolor{resultshade1}.459 & \cellcolor{resultshade2}.993 & \cellcolor{resultshade3}1.148 & \cellcolor{resultshade2}.187 & \cellcolor{resultshade3}.609 & \cellcolor{resultshade3}.568 & \cellcolor{resultshade3}1.202 & \cellcolor{resultshade4}\secondval{1.658} & \cellcolor{resultshade3}.759 & \cellcolor{resultshade2}.434 & \cellcolor{resultshade1}.266 & \cellcolor{resultshade1}.793 & \cellcolor{resultshade1}.010 \\
\addlinespace[2pt]
RoBERTa & \cellcolor{resultshade1}.709 & \cellcolor{resultshade1}.481 & \cellcolor{resultshade5}\bestval{.748} & \cellcolor{resultshade5}\bestval{.944} & \cellcolor{resultshade4}\secondval{.243} & \cellcolor{resultshade4}\secondval{.657} & \cellcolor{resultshade4}\secondval{.631} & \cellcolor{resultshade4}\secondval{1.162} & \cellcolor{resultshade4}\secondval{1.658} & \cellcolor{resultshade4}\secondval{.786} & \cellcolor{resultshade2}.436 & \cellcolor{resultshade2}.431 & \cellcolor{resultshade3}.853 & \cellcolor{resultshade3}.609 \\
DeBERTa & \cellcolor{resultshade2}.716 & \cellcolor{resultshade3}.501 & \cellcolor{resultshade4}\secondval{.749} & \cellcolor{resultshade4}\secondval{.946} & \cellcolor{resultshade5}\bestval{.245} & \cellcolor{resultshade5}\bestval{.662} & \cellcolor{resultshade5}\bestval{.637} & \cellcolor{resultshade5}\bestval{1.121} & \cellcolor{resultshade5}\bestval{1.623} & \cellcolor{resultshade5}\bestval{.789} & \cellcolor{resultshade3}.438 & \cellcolor{resultshade3}.436 & \cellcolor{resultshade2}.841 & \cellcolor{resultshade2}.596 \\
\addlinespace[2pt]
Qwen3-4B & \cellcolor{resultshade2}.715 & \cellcolor{resultshade3}.539 & \cellcolor{resultshade1}1.392 & \cellcolor{resultshade1}1.558 & \cellcolor{resultshade1}.026 & \cellcolor{resultshade1}.322 & \cellcolor{resultshade1}.279 & \cellcolor{resultshade1}2.504 & \cellcolor{resultshade1}3.441 & \cellcolor{resultshade1}.050 & \cellcolor{resultshade1}.363 & \cellcolor{resultshade1}.272 & \cellcolor{resultshade1}.793 & \cellcolor{resultshade2}.491 \\
Qwen3-4B LoRA & \cellcolor{resultshade3}.719 & \cellcolor{resultshade2}.491 & \cellcolor{resultshade4}.782 & \cellcolor{resultshade4}1.017 & \cellcolor{resultshade3}.219 & \cellcolor{resultshade4}.656 & \cellcolor{resultshade4}.621 & \cellcolor{resultshade4}1.183 & \cellcolor{resultshade3}1.772 & \cellcolor{resultshade4}.767 & \cellcolor{resultshade3}.461 & \cellcolor{resultshade3}.461 & \cellcolor{resultshade3}.848 & \cellcolor{resultshade3}.634 \\
\addlinespace[2pt]
Qwen3.6 & \cellcolor{resultshade1}.566 & \cellcolor{resultshade1}.434 & \cellcolor{resultshade1}1.186 & \cellcolor{resultshade1}1.629 & \cellcolor{resultshade2}.206 & \cellcolor{resultshade1}.419 & \cellcolor{resultshade1}.283 & \cellcolor{resultshade1}2.282 & \cellcolor{resultshade1}2.775 & \cellcolor{resultshade1}-.172 & \cellcolor{resultshade1}.421 & \cellcolor{resultshade2}.421 & \cellcolor{resultshade2}.846 & \cellcolor{resultshade1}.438 \\
GPT-6-Astra & \cellcolor{resultshade4}\secondval{.747} & \cellcolor{resultshade4}\secondval{.562} & \cellcolor{resultshade2}1.098 & \cellcolor{resultshade1}1.421 & \cellcolor{resultshade3}.211 & \cellcolor{resultshade1}.319 & \cellcolor{resultshade2}.309 & \cellcolor{resultshade2}2.024 & \cellcolor{resultshade2}2.662 & \cellcolor{resultshade2}.333 & \cellcolor{resultshade5}\bestval{.635} & \cellcolor{resultshade4}\secondval{.533} & \cellcolor{resultshade5}\bestval{.919} & \cellcolor{resultshade5}\bestval{.805} \\
Gemini-3.1 & \cellcolor{resultshade5}\bestval{.795} & \cellcolor{resultshade5}\bestval{.587} & \cellcolor{resultshade3}.814 & \cellcolor{resultshade2}1.231 & \cellcolor{resultshade5}\bestval{.245} & \cellcolor{resultshade2}.446 & \cellcolor{resultshade2}.390 & \cellcolor{resultshade2}2.045 & \cellcolor{resultshade2}2.564 & \cellcolor{resultshade2}.252 & \cellcolor{resultshade4}\secondval{.631} & \cellcolor{resultshade4}.513 & \cellcolor{resultshade4}.894 & \cellcolor{resultshade4}\secondval{.732} \\
DeepSeek-V4.1 & \cellcolor{resultshade1}.688 & \cellcolor{resultshade4}.547 & \cellcolor{resultshade1}1.125 & \cellcolor{resultshade2}1.411 & \cellcolor{resultshade4}.228 & \cellcolor{resultshade2}.438 & \cellcolor{resultshade1}.278 & \cellcolor{resultshade1}2.063 & \cellcolor{resultshade3}2.530 & \cellcolor{resultshade1}-.048 & \cellcolor{resultshade4}.576 & \cellcolor{resultshade5}\bestval{.542} & \cellcolor{resultshade4}\secondval{.899} & \cellcolor{resultshade4}.710 \\
\bottomrule
\end{tabular*}
\end{table*}

\subsection{Management: Policy Outcomes and Temporal Dynamics}
\label{sec:policy-events}

Content and interaction tasks characterize individual contributions and responses. Platform management introduces another level of uncertainty: how participation, visibility, and audience activity evolve when platform rules change. Policy decisions can alter who is eligible to contribute, how annotations are reviewed or displayed, and what incentives contributors receive. Anticipating these changes requires reasoning about aggregate outcomes and their evolution over time~\cite{lloyd2026beyond}.

We ground this evaluation in five documented Weibo Annotations policy events, covering eligibility, distribution, rule documentation, incentives, and institutional participation:

\begin{itemize}[leftmargin=*]
    \item \textbf{P1. Eligibility, interaction, and display (Nov.\ 7, 2023).} Expanded eligibility; introduced card-level interactions, multiple displayed annotations, and a content-copying category.\footnote{\url{https://weibo.com/1934183965/NrtVG5zfG}}
    
    \item \textbf{P2. Feedback and distribution (Dec.\ 20, 2023).} Added approval feedback and increased annotation exposure.\footnote{\url{https://weibo.com/1934183965/Ny18nyzW2}}
    
    \item \textbf{P3. Rule documentation (Apr.\ 23, 2024).} Updated descriptions of annotation types, review, display, and incentives; implementation timing remains unverified.\footnote{\url{https://kefu.weibo.com/faqdetail?id=21748}}
    
    \item \textbf{P4. Revenue Incentive Model (May 7, 2025).} Introduced CNY 5--50 rewards for eligible approved annotations and potential advertising revenue.\footnote{\url{https://www.sina.cn/news/detail/5163349935653372.html}}
    
    \item \textbf{P5. Enterprise responses and post reuse (Oct.\ 14, 2025).} Added labeled enterprise responses and expanded reuse of existing posts.\footnote{\url{https://weibo.com/1934183965/Q94bsiZr7}}
\end{itemize}

To evaluate activity across these events, we define four daily outcomes capturing different aspects of platform functioning: \emph{submissions} (contribution volume), \emph{active contributors} (participation breadth), \emph{displayed annotations} (public visibility), and \emph{engagement index} (audience response)~\cite{ren2014community,arjmandi2025sustainability,weibofaq,weiboChiStudy}. The engagement index averages $\log(1+e_i)$ across annotations submitted on each day, where $e_i$ is archived engagement. Display counts use submission-date cohorts and collection-time status. As the records are observational, the tasks evaluate policy-associated activity without attributing observed changes causally to the policies.

\vspace{0.2em}
\textbf{T6: Policy outcome QA.}
T6 predicts increases, stability, or decreases in platform activity from policy text and pre-policy data. Counterfactual baselines use Bayesian dynamic regression fitted to pre-policy data, with a linear trend, weekday effects, AR(1) errors, and square-root-transformed counts. Posterior simulations provide 90\% prediction intervals. Labels compare observed and counterfactual means over days 1--14 and 29--90 using a $\pm20\%$ stability band. Four outcomes across five events yield 40 questions relevant to platform planning~\cite{weiboChiStudy} (Appendix~\ref{sec:policy-description}).

\vspace{0.2em}
\textbf{T7: Policy trajectory simulation.}
Predicting the direction of change does not establish whether a model can anticipate how activity unfolds over time. T7 addresses this distinction through LLM-guided calibration of a shared agent-based model (ABM), which forecasts daily policy-associated activity~\cite{ren2014community,muric2020socialsim,capstick2024autoelicit}. Calibration uses policy text and the first 10 post-policy days for P1--P2 or 30 days for P3--P5. Model parameters are subsequently frozen for 20- or 60-day forecasts, evaluated using MAE, RMSE, dynamic time warping (DTW), and Pearson $r$ (Appendix~\ref{sec:abm-protocol}).

T6 and T7 thus assess complementary levels of policy reasoning. A correct prediction of an overall increase or decrease need not imply an accurate forecast of the timing, magnitude, or fluctuations of platform activity. This distinction matters when simulation is used to inform review capacity, contributor support, and other operational planning decisions.

\section{Model Capabilities and Blind Spots}
\label{sec:experiments}

\dataset{} provides a shared empirical setting for evaluating how models assess corrective contributions (Section~\ref{sec:contribution-results}), interpret audience responses (Section~\ref{sec:public-interaction-results}), and anticipate policy-associated activity (Section~\ref{sec:policy-results}). We compare model families across seven tasks and examine context-specific blind spots associated with contribution types, comment stances, and policy trajectories.

\subsection{Evaluation Protocol}

We evaluate ten models on T1--T5 and four LLMs on T6--T7, with 12 human forecasters providing a T6 reference. Models span three families: \textbf{ML} methods (XGBoost~\cite{chen2016xgboost}, Random Forest~\cite{breiman2001forests}), \textbf{LM} encoders (Chinese RoBERTa~\cite{cui2021pretraining}, DeBERTa~\cite{he2021deberta}), and \textbf{LLMs} (Qwen3-4B~\cite{yang2025qwen3} with and without LoRA~\cite{hu2022lora}, Qwen3.6-35B-A3B~\cite{qwen36}, GPT-6-Astra~\cite{gpt6astra}, Gemini-3.1-Pro~\cite{gemini31}, and DeepSeek-V4.1-Flash~\cite{deepseek41}).

Supervised models use task-specific training, validation-based selection, and three seeds; LLM evaluations use three runs. For T7, three parameter proposals are combined before calibration, yielding one deterministic ABM forecast per model and policy. Appendix~\ref{sec:repro} provides implementation details.

Models share test instances within each task, with 1,822 notes shared across T1--T3 (Table~\ref{tab:main-results}). We report accuracy and Macro-F1 for classification and policy QA, Toxic-F1 for toxicity~\cite{schafer2024hierarchical}, and MAE, RMSE, and correlation for quality and engagement regression. T3 regression uses $\ln(1+\text{engagement})$; simulation metrics follow Section~\ref{sec:benchmark}. Appendix Table~\ref{tab:model-variance} reports run variability.

\begin{figure*}[t]
\centering\includegraphics[width=\textwidth]{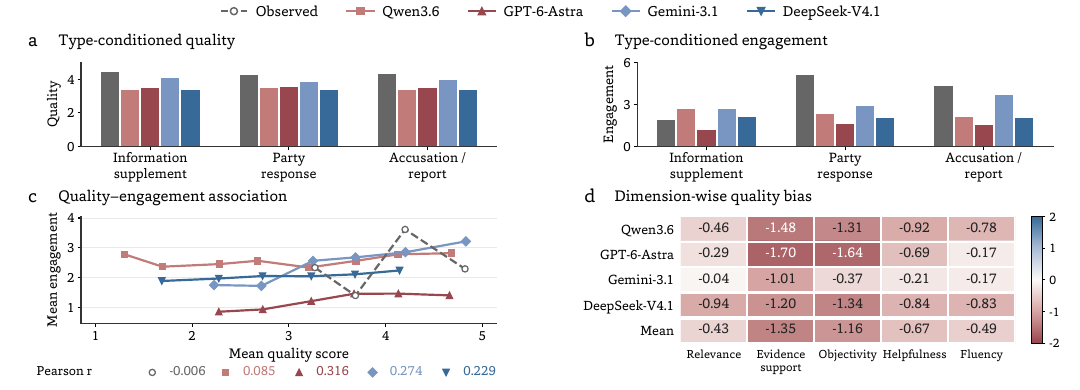}
\caption{Type-conditioned predictions, quality--engagement associations, and systematic biases in zero-shot LLMs.
(a) Mean quality scores and (b) mean log engagement across annotation types, comparing model predictions with observed references.
(c) Quality--engagement associations in observed data and model predictions, with grouped trends and Pearson correlations ($r$).
(d) Mean signed quality errors (prediction minus human reference) across five dimensions: relevance, evidence support, objectivity, helpfulness, and fluency.}
\Description{Panels (a,b) compare observed and predicted mean quality and engagement across information supplements, party responses, and accusations or reports. Panel (c) shows binned quality--engagement associations. Pearson correlations are -0.006 for observed data, 0.085 for Qwen3.6, 0.316 for GPT-6-Astra, 0.274 for Gemini-3.1, and 0.229 for DeepSeek-V4.1. Panel (d) shows signed prediction errors across five quality dimensions, with a final row averaging the four models.}
\label{fig:capability-diagnostic}
\end{figure*}


\subsection{Content: Quality and Attention (T1--T3)}
\label{sec:contribution-results}

Table~\ref{tab:main-results} reports T1--T3 performance, with Gemini-3.1 leading type classification (Macro-F1 0.587). We examine the reliability of quality judgments and their relationship with audience engagement.

\textbf{Low rating errors mask weak agreement and systematic quality bias.}
Fine-tuned RoBERTa and DeBERTa achieve the lowest T2 MAEs (0.748/0.749), yet correlate weakly with human ratings ($r=0.243/0.245$; Table~\ref{tab:main-results}). LoRA reduces Qwen3-4B's MAE from 1.392 to 0.782, but its correlation remains at 0.219. All four zero-shot LLMs underrate quality across contribution types (Figure~\ref{fig:capability-diagnostic}(a)), particularly in evidence support and objectivity (Figure~\ref{fig:capability-diagnostic}(d)), with biases reaching $-1.70$ and $-1.64$ for GPT-6-Astra. Systematic underestimation in these dimensions may disadvantage well-supported contributions during automated review.

\textbf{LLMs compress engagement differences and overstate their association with quality.}
Engagement predictions compress observed differences across contribution types, particularly for party responses and accusations/reports (Figure~\ref{fig:capability-diagnostic}(b)). Notably, reference quality varies little across contribution types, while observed engagement differs substantially (Figure~\ref{fig:capability-diagnostic}(a,b)). Quality and engagement are nearly uncorrelated in matched references ($r=-0.006$), whereas GPT-6-Astra, Gemini-3.1, and DeepSeek-V4.1 predict stronger positive associations ($r=0.316$, $0.274$, and $0.229$; Figure~\ref{fig:capability-diagnostic}(c)). Qwen3.6 is closer ($r=0.085$), despite substantial prediction errors. Task-specific supervision improves engagement prediction: DeBERTa achieves 0.637 Macro-F1 versus 0.390 for the best zero-shot LLM, while LoRA raises Qwen3-4B from 0.279 to 0.621 (Table~\ref{tab:main-results}). These mismatches caution against using predicted corrective quality as a proxy for audience attention.

\begin{table}[!b]
\caption{\textbf{Policy outcome QA (T6).}
Accuracy and Macro-F1 on 40 targets spanning five policy events, four outcomes, and two horizons: short-term (days 1--14) and long-term (days 29--90).
Results average three LLM runs or 12 human forecasters.
Five purple shades indicate within-column ranks (darker is better); bold and underlined values mark best and second-best, respectively, including ties.}
\label{tab:policy-outcome-qa}
\centering
\normalfont\small
\setlength{\tabcolsep}{3pt}\renewcommand{\arraystretch}{1.08}
\begin{tabular*}{\columnwidth}{@{\extracolsep{\fill}}lrrrrrr@{}}
\toprule
& \multicolumn{2}{c}{\textsc{Short term}} & \multicolumn{2}{c}{\textsc{Long term}} & \multicolumn{2}{c}{\textsc{Overall}} \\
\cmidrule(lr){2-3}\cmidrule(lr){4-5}\cmidrule(l){6-7}
Model & Acc.$\uparrow$ & F1$\uparrow$ & Acc.$\uparrow$ & F1$\uparrow$ & Acc.$\uparrow$ & F1$\uparrow$ \\
\midrule
Qwen3.6 & \cellcolor{resultshade2}.533 & \cellcolor{resultshade3}.429 & \cellcolor{resultshade1}.250 & \cellcolor{resultshade1}.281 & \cellcolor{resultshade1}.392 & \cellcolor{resultshade1}.380 \\
GPT-6-Astra & \cellcolor{resultshade5}\bestval{.700} & \cellcolor{resultshade5}\bestval{.617} & \cellcolor{resultshade5}\bestval{.633} & \cellcolor{resultshade5}\bestval{.592} & \cellcolor{resultshade5}\bestval{.667} & \cellcolor{resultshade5}\bestval{.622} \\
Gemini-3.1 & \cellcolor{resultshade2}.533 & \cellcolor{resultshade1}.418 & \cellcolor{resultshade4}\secondval{.533} & \cellcolor{resultshade4}\secondval{.442} & \cellcolor{resultshade4}\secondval{.533} & \cellcolor{resultshade3}.434 \\
DeepSeek-V4.1 & \cellcolor{resultshade3}\secondval{.567} & \cellcolor{resultshade2}.423 & \cellcolor{resultshade2}.467 & \cellcolor{resultshade2}.344 & \cellcolor{resultshade3}.517 & \cellcolor{resultshade2}.385 \\
Human & \cellcolor{resultshade1}.512 & \cellcolor{resultshade4}\secondval{.435} & \cellcolor{resultshade3}.487 & \cellcolor{resultshade3}.437 & \cellcolor{resultshade2}.500 & \cellcolor{resultshade4}\secondval{.439} \\
\bottomrule
\end{tabular*}
\end{table}
\begin{figure}[!t]
\centering\includegraphics[width=\columnwidth]{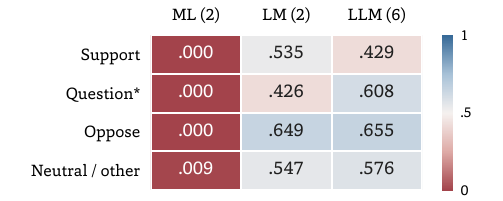}
\caption{\textbf{Toxic-comment recall across reader stances and model families (T4--T5).}
Each cell reports toxic-class recall within a reference stance, averaged over three runs per model and then across models within each family (ML: 2; LM: 2; LLM: 6).
Invalid predictions count as misses. Colors share a 0--1 scale, from red (lower recall) to blue (higher recall).}
\Description{A single-column heatmap reports non-toxic and toxic Recall and F1 for support, question, oppose, and neutral or other stances. The two metric groups share one stance axis and one right-side color scale. Toxic Recall and F1 are lower than their non-toxic counterparts across all four stances; supportive toxic comments have the lowest average recall. All sixteen cells have exact numerical annotations.}
\label{fig:toxicity-diagnostic}
\end{figure}
\begin{figure*}[!t]
\centering\includegraphics[width=0.9\textwidth]{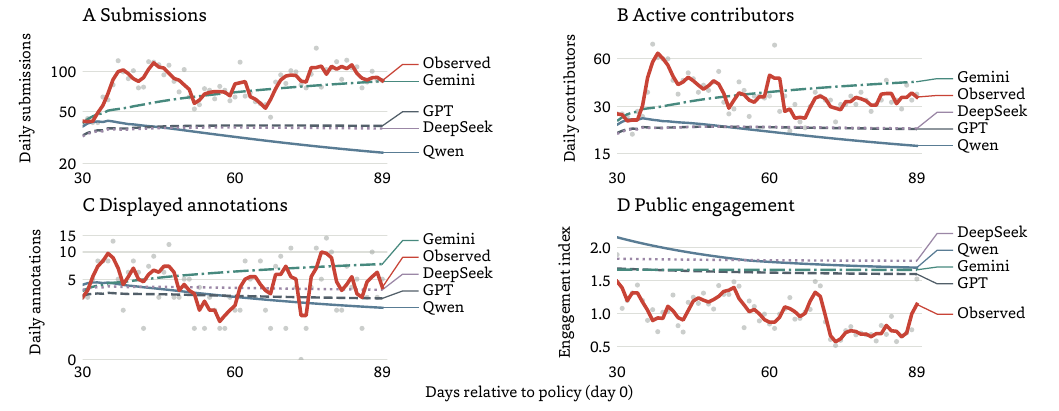}
\caption{\textbf{Observed activity and T7 forecasts after P4 (Revenue Incentive Model).}
Gray dots show held-out daily observations (days 30--89), red curves their centered three-day means, and labeled curves deterministic ABM forecasts with frozen LLM-guided parameters.
Count ticks retain original units. Appendix~\ref{sec:abm-protocol} details the metrics; Figure~\ref{fig:policy-trajectories} covers all five policies.}
\Description{Four panels compare P4 submissions, active contributors, displayed annotations, and engagement with deterministic ABM point forecasts under Qwen, GPT, Gemini, and DeepSeek parameter modifications. Curves are labeled directly at their endpoints. Daily observations fluctuate around comparatively smooth point forecasts, and observed engagement remains below all four forecast paths. Raw points remain visible and outcome-specific axes include every plotted value.}
\label{fig:simulation-fidelity}
\end{figure*}
\begin{table}[!t]
\caption{\textbf{Policy trajectory simulation (T7).}
Frozen LLM-guided ABM forecasts are compared with observed daily activity, with metrics averaged equally across five policies per outcome.
MAE/RMSE/DTW retain outcome units ($\downarrow$); Pearson $r$ measures temporal agreement ($\uparrow$).
Darker purple indicates better within-outcome ranks; bold/underlining mark best/second-best, including ties.}
\label{tab:policy-simulation}
\centering\normalfont\small
\setlength{\tabcolsep}{3pt}\renewcommand{\arraystretch}{1.08}
\begin{tabular*}{\columnwidth}{@{\extracolsep{\fill}}lrrrr@{}}
\toprule
Model & MAE$\downarrow$ & RMSE$\downarrow$ & DTW$\downarrow$ & $r\uparrow$ \\
\midrule
\multicolumn{5}{l}{\textsc{Submissions}} \\
Qwen3.6 & \cellcolor{resultshade3}\secondval{19.678} & \cellcolor{resultshade1}23.803 & \cellcolor{resultshade1}154.821 & \cellcolor{resultshade1}$-.056$ \\
GPT-6-Astra & \cellcolor{resultshade1}20.256 & \cellcolor{resultshade2}23.656 & \cellcolor{resultshade3}\secondval{152.654} & \cellcolor{resultshade3}\secondval{.129} \\
Gemini-3.1 & \cellcolor{resultshade5}\bestval{15.622} & \cellcolor{resultshade5}\bestval{19.423} & \cellcolor{resultshade5}\bestval{108.981} & \cellcolor{resultshade5}\bestval{.163} \\
DeepSeek-V4.1 & \cellcolor{resultshade2}20.145 & \cellcolor{resultshade3}\secondval{23.650} & \cellcolor{resultshade2}154.033 & \cellcolor{resultshade2}.116 \\
\midrule
\multicolumn{5}{l}{\textsc{Active contributors}} \\
Qwen3.6 & \cellcolor{resultshade5}\bestval{8.350} & \cellcolor{resultshade5}\bestval{10.232} & \cellcolor{resultshade3}\secondval{63.375} & \cellcolor{resultshade5}\bestval{.125} \\
GPT-6-Astra & \cellcolor{resultshade1}9.929 & \cellcolor{resultshade1}11.962 & \cellcolor{resultshade1}71.065 & \cellcolor{resultshade3}\secondval{.102} \\
Gemini-3.1 & \cellcolor{resultshade2}9.546 & \cellcolor{resultshade2}11.547 & \cellcolor{resultshade5}\bestval{62.206} & \cellcolor{resultshade1}.011 \\
DeepSeek-V4.1 & \cellcolor{resultshade3}\secondval{9.394} & \cellcolor{resultshade3}\secondval{11.311} & \cellcolor{resultshade2}69.517 & \cellcolor{resultshade2}.083 \\
\midrule
\multicolumn{5}{l}{\textsc{Displayed annotations}} \\
Qwen3.6 & \cellcolor{resultshade5}\bestval{2.218} & \cellcolor{resultshade5}\bestval{2.884} & \cellcolor{resultshade5}\bestval{17.969} & \cellcolor{resultshade2}$-.078$ \\
GPT-6-Astra & \cellcolor{resultshade2}2.442 & \cellcolor{resultshade2}3.112 & \cellcolor{resultshade3}\secondval{18.920} & \cellcolor{resultshade5}\bestval{$-.066$} \\
Gemini-3.1 & \cellcolor{resultshade1}2.602 & \cellcolor{resultshade1}3.191 & \cellcolor{resultshade2}18.924 & \cellcolor{resultshade1}$-.101$ \\
DeepSeek-V4.1 & \cellcolor{resultshade3}\secondval{2.330} & \cellcolor{resultshade3}\secondval{2.973} & \cellcolor{resultshade1}18.929 & \cellcolor{resultshade3}\secondval{$-.075$} \\
\midrule
\multicolumn{5}{l}{\textsc{Engagement index}} \\
Qwen3.6 & \cellcolor{resultshade1}.647 & \cellcolor{resultshade1}.772 & \cellcolor{resultshade2}4.392 & \cellcolor{resultshade2}.072 \\
GPT-6-Astra & \cellcolor{resultshade3}\secondval{.602} & \cellcolor{resultshade3}\secondval{.745} & \cellcolor{resultshade3}\secondval{4.351} & \cellcolor{resultshade1}$-.062$ \\
Gemini-3.1 & \cellcolor{resultshade5}\bestval{.595} & \cellcolor{resultshade5}\bestval{.735} & \cellcolor{resultshade5}\bestval{4.296} & \cellcolor{resultshade5}\bestval{.102} \\
DeepSeek-V4.1 & \cellcolor{resultshade2}.621 & \cellcolor{resultshade2}.749 & \cellcolor{resultshade1}4.432 & \cellcolor{resultshade3}\secondval{.077} \\
\bottomrule
\end{tabular*}
\end{table}

\subsection{Interaction: Stance and Safety (T4--T5)}
\label{sec:public-interaction-results}
Reader stance and interaction safety capture distinct aspects of public feedback: whether a comment supports a correction and whether it contains harmful content. We examine aggregate performance on T4--T5 (Table~\ref{tab:main-results}) and stance-conditioned toxicity detection (Figure~\ref{fig:toxicity-diagnostic}).

\textbf{Aggregate accuracy can conceal severe toxicity detection failures.}
ML baselines predict non-toxic for over 99\% of comments, achieving approximately 0.793 accuracy but only 0.010 Toxic-F1 (Table~\ref{tab:main-results}). Language models perform substantially better: GPT-6-Astra achieves the highest Toxic-F1 (0.805), while LoRA improves Qwen3-4B from 0.491 to 0.634. Yet stronger aggregate detection does not establish which harmful interactions models reliably recognize.

\textbf{Supportive toxicity is a persistent blind spot for LLMs.}
Figure~\ref{fig:toxicity-diagnostic} shows that LLMs recall only 42.9\% of toxic supportive comments, compared with 60.8\% of questioning and 65.5\% of opposing comments. Fine-tuned LM encoders reach 53.5\% recall on supportive toxicity, while ML baselines detect virtually none. Across all ten models, supportive toxicity has the lowest recall (36.4\%; $n=67$). Although this subgroup is relatively small, the pattern shows how endorsement of a correction can coexist with abuse that moderation models overlook.



\subsection{Management: Directional Accuracy and Temporal Fidelity (T6--T7)}

\label{sec:policy-results}

Policy-associated activity can be evaluated at two resolutions: predicting outcome directions over a policy horizon (T6) and reproducing daily activity trajectories (T7). We assess directional predictions (Table~\ref{tab:policy-outcome-qa}) and simulation fidelity (Table~\ref{tab:policy-simulation}; Figure~\ref{fig:simulation-fidelity}).

\textbf{Directional accuracy offers limited assurance of temporal fidelity.}
GPT-6-Astra achieves 0.667 accuracy and 0.622 Macro-F1 on T6, compared with mean human accuracy of 0.500 (Table~\ref{tab:policy-outcome-qa}). Yet LLM-guided ABM forecasts on T7 correlate weakly with observed daily activity, with mean Pearson $r$ ranging from $-0.101$ to $0.163$ across model--outcome pairs (Table~\ref{tab:policy-simulation}). This contrast shows that identifying broad outcome directions provides limited information about the timing and variation of platform responses.

\textbf{Low forecasting errors can conceal incorrect temporal dynamics.}
Qwen3.6 achieves the lowest displayed-annotation MAE (2.218), yet its temporal correlation remains negative ($r=-0.078$). Similarly, Gemini-3.1 attains the lowest submission MAE (15.622) but only $r=0.163$ (Table~\ref{tab:policy-simulation}). Figure~\ref{fig:simulation-fidelity} reveals comparatively smooth deterministic forecasts that miss observed peaks and reversals. Although some local trends are recovered, such as active contributors after P1 ($r=0.357$; Appendix Figure~\ref{fig:policy-trajectories}), agreement remains weak overall. These discrepancies highlight the need to evaluate temporal structure alongside aggregate forecasting error.


\section{Discussion}

Our findings reveal how unreliable local judgments can misguide platform decisions: biased quality ratings affect review, popularity obscures valuable corrections, supportive feedback conceals abuse, and coarse forecasts miss operational fluctuations.

\vspace{0.2em}
\textbf{Evidence-grounded review.}
Weak agreement with human ratings and type-specific errors make automated acceptance thresholds risky. Misjudging evidence support or objectivity could exclude valuable contributions and discourage continued participation~\cite{weiboChiStudy,arjmandi2025sustainability}. AI assessments should identify specific evidentiary gaps and provide traceable support, enabling reviewers to verify judgments and contributors to revise their notes before publication.

\vspace{0.2em}
\textbf{Identifying overlooked corrections.}
Quality correlates weakly with engagement ($\rho=0.068$), raising concerns that popularity-driven ranking may overlook valuable notes. High-quality, low-engagement contributions warrant visibility audits, particularly across niche topics and smaller contributor groups. These audits should distinguish limited exposure from weak audience interest, ideally using impression-level data, since engagement alone cannot establish whether readers encountered a correction~\cite{weiboChiStudy}.

\vspace{0.2em}
\textbf{Moderating harm across stances.}
Models frequently miss toxic endorsement, exposing a blind spot in moderation of supportive feedback. Stance-conditioned toxicity audits can reveal such failures, while independent assessments of stance and harm can help flag abusive support without suppressing constructive disagreement. This distinction matters when hostile feedback increases the social cost of correction and may deter future participation~\cite{chuai2025outrage,smirnov2023toxic}.

\vspace{0.2em}
\textbf{Testing policy forecasts against reality.}
Directional accuracy offers limited guidance for operational planning when forecasts miss daily peaks and reversals. Platforms should evaluate temporal fidelity alongside aggregate errors and recalibrate against observed submissions, participation, display, and engagement~\cite{weiboChiStudy,hut2026simulation}. Forecast errors also reflect ABM assumptions and calibration, while observational policy associations cannot establish causal effects. Forecasts should therefore inform monitoring and scenario analysis, with operational decisions grounded in observed activity.

\vspace{0.2em}
\textbf{Making human--AI handoffs accountable.}
Reliable assistance requires targeted human oversight in evidence verification, visibility review, harm moderation, and policy monitoring. Linking evidence, review outcomes, display records, and reader feedback could support auditable handoffs, contributor revisions, and operator overrides~\cite{lloyd2026beyond,amershi2019guidelines}. Prior work shows that human challenges to AI-generated notes can improve helpfulness~\cite{de2026feedback}; whether integrated workflows improve corrective reach and interaction safety remains untested.

\section{Conclusion}

A correction's quality tells only part of its story. Its reach and reception depend on audience behavior and platform mechanisms. We introduce \dataset{}, a seven-task benchmark of correction context understanding spanning content, interaction, and management, grounded in linked Weibo Annotations data. Our evaluation reveals that quality judgments align weakly with human assessments, quality correlates weakly with engagement, supportive toxicity often goes undetected, and policy forecasts miss daily dynamics despite predicting broad directions. Cross-group analyses expose type- and stance-specific failures masked by aggregate scores. These findings motivate evidence-grounded review, visibility and safety audits, and policy forecasts validated against observed activity, with human oversight guiding platform decisions. Reliable fact-checking AI must understand the note, the conversation around it, and the platform in which both unfold.

\bibliographystyle{ACM-Reference-Format}
\bibliography{references}

@article{hellofresh,
 author={Tim Franzmeyer and Aleksandar Shtedritski and Samuel Albanie and Philip Torr and Jo{\~a}o F. Henriques and Jakob N. Foerster},
 title={{HelloFresh}: {LLM} Evaluations on Streams of Real-World Human Editorial Actions across {X Community Notes} and Wikipedia edits},
 journal={arXiv preprint arXiv:2406.03428}, year={2024}}

@inproceedings{communitynotes,
 author={Rui Xing and Preslav Nakov and Timothy Baldwin and Jey Han Lau},
 title={{COMMUNITYNOTES}: A Dataset for Exploring the Helpfulness of Fact-Checking Explanations},
 booktitle={Findings of the Association for Computational Linguistics: EACL 2026},
 year={2026}, pages={1390--1411}, publisher={Association for Computational Linguistics}}

@article{healthnotes,
 author={Jiaying Wu and Zihang Fu and Haonan Wang and Fanxiao Li and Jiafeng Guo and Preslav Nakov and Min-Yen Kan},
 title={Beyond the Crowd: {LLM}-Augmented Community Notes for Governing Health Misinformation},
 journal={arXiv preprint arXiv:2510.11423},year={2026},note={Version 4}}

@article{polbench,
 author={Sahajpreet Singh and Kokil Jaidka and Min-Yen Kan},
 title={{GitSearch}: Enhancing Community Notes Generation with Gap-Informed Targeted Search},
 journal={arXiv preprint arXiv:2602.08945},year={2026}}

@article{comrate,
 author={Changxi Wen and Shuning Zhang and Bohao Chu and Yuwei Chuai and Hui Wang and Dai Shi and Xin Yi and Hewu Li},
 title={Towards Multi-Agent-Simulation-Based Community Note Evaluation},
 journal={arXiv preprint arXiv:2606.18268},year={2026}}

@article{supernotes,
 author={Soham De and Michiel A. Bakker and Jay Baxter and Martin Saveski},
 title={Supernotes: Driving Consensus in Crowd-Sourced Fact-Checking},
 journal={arXiv preprint arXiv:2411.06116},year={2024}}

@inproceedings{rumoureval,
 author={Genevieve Gorrell and Elena Kochkina and Maria Liakata and Ahmet Aker and Arkaitz Zubiaga and Kalina Bontcheva and Leon Derczynski},
 title={{SemEval}-2019 Task 7: {RumourEval}, Determining Rumour Veracity and Support for Rumours},
 booktitle={Proceedings of the 13th International Workshop on Semantic Evaluation},
 year={2019},pages={845--854},publisher={Association for Computational Linguistics}}

@misc{fouryearcn,
 author={Saeedeh Mohammadi and Narges Chinichian and Hannah Doyal and Kristina Skutilova and Bertani Anna and Cui Hao and Michele d'Errico and Siobhan Grayson and Taha Yasseri},
 title={From Birdwatch to Community Notes, from Twitter to X: Four Years of Community-Based Content Moderation},
 publisher={Zenodo}, year={2026}, doi={10.5281/zenodo.20591253}}

@misc{gpt6astra,
 author={{OpenAI}}, title={{GPT-6 Astra} Model Documentation}, year={2026},
 url={https://developers.openai.com/api/docs/models/gpt-6-astra}, note={Accessed 2026-09-27}}

@misc{gemini31,
 author={{Google}}, title={Gemini 3.1 Pro Preview}, year={2026},
 url={https://ai.google.dev/gemini-api/docs/models/gemini-3.1-pro-preview}, note={Accessed 2026-09-27}}

@misc{deepseek41,
 author={{DeepSeek-AI}}, title={DeepSeek-V4.1-Flash: Smarter, Faster, More Efficient}, year={2026},
 url={https://api-docs.deepseek.com/news/news260910/}, note={Accessed 2026-09-27}}

@misc{weibolaunch,
 author={{Weibo}}, title={Announcement of the Weibo Annotations Feature}, year={2023},
 howpublished={Weibo article}, url={https://www.weibo.com/ttarticle/p/show?id=2309404930349917667993}, note={Published 2023-08-02; accessed 2026-09-27}}

@misc{weibofaq,
 author={{Weibo Customer Service}}, title={Adding Annotations on Weibo}, year={2024},
 url={https://kefu.weibo.com/faqdetail?id=21748}, note={Updated 2024-04-23; accessed 2026-09-27}}

@article{weiboChiStudy,
 author={Lin, Chichen and Xie, Huaijin and Xie, Wenhan and Fan, Weijian and Han, Xiao and Jin, Yijie and Hao, Xiaojie and Wu, Jiaying},
 title={When Platforms Pay Users to Correct Misinformation: Revenue Incentives and Governance Capacity in {Weibo Annotations}},
 journal={arXiv preprint arXiv:2610.18668},
 year={2026}}

@inproceedings{volkovs2020engagement,
 author={Volkovs, Maksims and Cheng, Zhaoyue and Ravaut, Mathieu and Yang, Hojin and Shen, Kevin and Zhou, Jin Peng and Wong, Anson and Zuberi, Saba and Zhang, Ivan and Frosst, Nick and Ngo, Helen and Chen, Carol and Venkitesh, Bharat and Gou, Stephen and Gomez, Aidan N.},
 title={Predicting Twitter Engagement With Deep Language Models}, booktitle={Proceedings of the Recommender Systems Challenge 2020}, year={2020}}

@inproceedings{xu2025smtpd,
 author={Xu, Yijie and Zheng, Bolun and Zhu, Wei and Pan, Hangjia and Yao, Yuchen and Xu, Ning and Liu, Anan and Zhang, Quan and Yan, Chenggang},
 title={{SMTPD}: A New Benchmark for Temporal Prediction of Social Media Popularity}, booktitle={Proceedings of the IEEE/CVF Conference on Computer Vision and Pattern Recognition}, year={2025}}

@inproceedings{muric2020socialsim,
 author={Muric, Goran and Tregubov, Alexey and Blythe, Jim and Abeliuk, Andres and Choudhary, Divya and Lerman, Kristina and Ferrara, Emilio},
 title={Massive Cross-Platform Simulations of Online Social Networks}, booktitle={Proceedings of the 19th International Conference on Autonomous Agents and Multiagent Systems}, year={2020}, pages={895--903}, url={https://www.ifaamas.org/Proceedings/aamas2020/pdfs/p895.pdf}}

@article{ashokkumar2026experiments,
author={Ashwini Ashokkumar and Luke Hewitt and Isaias Ghezae and Robb Willer},
title={Large language models can predict the results of social science experiments},
journal={Nature}, volume={656}, pages={115--122}, year={2026}}

@inproceedings{hut2026simulation,
author={Stefan Hut and Lorenzo Masoero},
title={Can AI Agents Simulate A/B Test Outcomes? A Validation Framework for Agentic Experimentation},
booktitle={Conference on Language Modeling (COLM)},year={2026},
url={https://www.amazon.science/publications/can-ai-agents-simulate-a-b-test-outcomes-a-validation-framework-for-agentic-experimentation}}

@misc{weibop1,
 author={{Weibo Administrator}}, title={Weibo Annotations Feature Improvements and Expanded Eligibility}, year={2023},
 url={https://weibo.com/1934183965/NrtVG5zfG}, note={Announcement dated 2023-11-07; archived text and official customer-service link checked 2026-10-04}}

@inproceedings{chen2016xgboost,
 author={Tianqi Chen and Carlos Guestrin},
 title={{XGBoost}: A Scalable Tree Boosting System},
 booktitle={Proceedings of the 22nd ACM SIGKDD International Conference on Knowledge Discovery and Data Mining},
 year={2016}}

@article{breiman2001forests,
 author={Leo Breiman}, title={Random Forests},
 journal={Machine Learning}, volume={45}, pages={5--32}, year={2001}}

@article{ren2014community,
 author={Yuqing Ren and Robert E. Kraut},
 title={Agent-Based Modeling to Inform Online Community Design: Impact of Topical Breadth, Message Volume, and Discussion Moderation on Member Commitment and Contribution},
 journal={Human--Computer Interaction}, volume={29}, number={4}, pages={351--389},
 year={2014}}

@article{capstick2024autoelicit,
 author={Alexander Capstick and Rahul G. Krishnan and Payam Barnaghi},
 title={{AutoElicit}: Using Large Language Models for Expert Prior Elicitation in Predictive Modelling},
 journal={arXiv preprint arXiv:2411.17284}, year={2024}}

@inproceedings{liu2023geval,
 author={Yang Liu and Dan Iter and Yichong Xu and Shuohang Wang and Ruochen Xu and Chenguang Zhu},
 title={{G-Eval}: {NLG} Evaluation using {GPT-4} with Better Human Alignment},
 booktitle={Proceedings of the 2023 Conference on Empirical Methods in Natural Language Processing},
 year={2023}, pages={2511--2522}}

@inproceedings{rottger2021hatecheck,
 author={Paul R{\"o}ttger and Bertie Vidgen and Dong Nguyen and Zeerak Waseem and Helen Margetts and Janet Pierrehumbert},
 title={{HateCheck}: Functional Tests for Hate Speech Detection Models},
 booktitle={Proceedings of the 59th Annual Meeting of the Association for Computational Linguistics and the 11th International Joint Conference on Natural Language Processing (Volume 1: Long Papers)},
 year={2021}, pages={41--58}}

@book{hyndman2021forecasting,
 author={Rob J. Hyndman and George Athanasopoulos},
 title={Forecasting: Principles and Practice}, edition={3},
 publisher={OTexts}, address={Melbourne, Australia}, year={2021},
 url={https://otexts.com/fpp3/}}

@inproceedings{amershi2019guidelines,
 author={Saleema Amershi and Dan Weld and Mihaela Vorvoreanu and Adam Fourney and Besmira Nushi and Penny Collisson and Jina Suh and Shamsi Iqbal and Paul N. Bennett and Kori Inkpen and Jaime Teevan and Ruth Kikin-Gil and Eric Horvitz},
 title={Guidelines for Human-{AI} Interaction},
 booktitle={Proceedings of the 2019 CHI Conference on Human Factors in Computing Systems},
 year={2019}, articleno={3}, numpages={13}}

@inproceedings{lloyd2026beyond,
 author={Travis Lloyd and Tung Nguyen and Karen Levy and Mor Naaman},
 title={Beyond Community Notes: A Framework for Understanding and Building Crowdsourced Context Systems for Social Media},
 booktitle={Proceedings of the 2026 CHI Conference on Human Factors in Computing Systems},
 year={2026}}

@article{smirnov2023toxic,
 author={Ivan Smirnov and Camelia Oprea and Markus Strohmaier},
 title={Toxic comments are associated with reduced activity of volunteer editors on {Wikipedia}},
 journal={PNAS Nexus}, volume={2}, number={12}, pages={pgad385}, year={2023}}

@article{li2025scaling,
 author={Haiwen Li and Soham De and Manon Revel and Andreas Haupt and Brad Miller and Keith Coleman and Jay Baxter and Martin Saveski and Michiel A. Bakker},
 title={Scaling Human Judgment in Community Notes with {LLMs}},
 journal={Journal of Online Trust and Safety}, volume={3}, number={1}, year={2025}}

@article{zhang2026canote,
 author={Shuning Zhang and Jingruo Chen and Yuwei Chuai and Dai Shi and Yifan Wang and Xin Yi and Hewu Li},
 title={{CANote}: Empowering Fact-checking Note Writing Through Scaffolded and Provenance-based Human-{AI} Collaboration},
 journal={arXiv preprint arXiv:2606.07101}, year={2026}}

@article{de2026feedback,
 author={Soham De and Isaac Slaughter and Jiawei Guo and Qiao-Yun Cheng and Jiayuan Yan and Sruti Banerjee and Martin Saveski},
 title={How Human Feedback Shapes {AI}-generated Community Notes},
 journal={arXiv preprint arXiv:2606.30905}, year={2026}}

@article{bouchaud2026polarizing,
 author={Paul Bouchaud and Pedro Ramaciotti},
 title={Community Notes undermoderate polarizing content by design creating risks in electoral processes},
 journal={Science Advances}, volume={12}, number={27}, pages={eaee6932}, year={2026}}

@inproceedings{chuai2025outrage,
 author={Yuwei Chuai and Anastasia Sergeeva and Gabriele Lenzini and Nicolas Pr{\"o}llochs},
 title={Community Fact-Checks Trigger Moral Outrage in Replies to Misleading Posts on Social Media},
 booktitle={Proceedings of the 2025 CHI Conference on Human Factors in Computing Systems},
 publisher={Association for Computing Machinery}, address={New York, NY, USA},
 articleno={962}, numpages={23}, year={2025}}

@article{arjmandi2025sustainability,
 author={Zahra Arjmandi-Lari and Alexios Mantzarlis and Tom Stafford},
 title={Threats to the sustainability of Community Notes on {X}},
 journal={arXiv preprint arXiv:2510.00650}, year={2025}}

@misc{xnotesdata,
  author = {{X}},
  title = {Community Notes: Downloading Data},
  howpublished = {Community Notes Guide},
  url = {https://communitynotes.x.com/guide/en/under-the-hood/download-data},
  urldate = {2026-10-15}
}

@article{vairetti2026media,
  title={Media Analytics via Machine Learning: Social Media Engagement Prediction for TV Channels},
  author={Vairetti, Carla and Gubbins, Juan Carlos and Maldonado, Sebasti{\'a}n},
  journal={Social Network Analysis and Mining},
  volume={16},
  pages={38},
  year={2026}
}

@inproceedings{srinivasan2013engaging,
  title={Will Your Facebook Post Be Engaging?},
  author={Srinivasan, Balaji Vasan and Natarajan, Anandhavelu and Sinha, Ritwik and Gupta, Vineet and Revankar, Shriram and Ravindran, Balaraman},
  booktitle={Proceedings of the First Workshop on User Engagement Optimization},
  pages={25--28},
  year={2013},
  publisher={ACM}
}

@inproceedings{wu2026beyond,
  title={Beyond the Crowd: LLM-Augmented Community Notes
         for Governing Health Misinformation},
  author={Wu, Jiaying and Fu, Zihang and Wang, Haonan
          and Li, Fanxiao and Guo, Jiafeng and
          Nakov, Preslav and Kan, Min-Yen},
  booktitle={Proceedings of the 64th Annual Meeting
             of the Association for Computational Linguistics},
  year={2026},
  pages={5152--5171}
}

@article{huang2025legitimacy,
  title={Content Moderation by LLM: From Accuracy to Legitimacy},
  author={Huang, Tao},
  journal={Artificial Intelligence Review},
  volume={58},
  pages={320},
  year={2025}
}

@inproceedings{cho2024moderators,
  title={Can Language Model Moderators Improve the Health
         of Online Discourse?},
  author={Cho, Hyundong and Liu, Shuai and Shi, Taiwei
          and Jain, Darpan and Rizk, Basem and
          Huang, Yuyang and Lu, Zixun and Wen, Nuan
          and Gratch, Jonathan and Ferrara, Emilio
          and May, Jonathan},
  booktitle={Proceedings of the 2024 Conference of the
             North American Chapter of the Association
             for Computational Linguistics:
             Human Language Technologies},
  pages={7478--7496},
  year={2024}
}

@inproceedings{huang2026policysim,
  title={PolicySim: An LLM-Based Agent Social Simulation
         Sandbox for Proactive Policy Optimization},
  author={Huang, Renhong and Tang, Ning and Xu, Jiarong
          and Cao, Yuxuan and Tu, Qingqian and
          Guo, Sheng and Zheng, Bo and Liu, Huiyuan
          and Yang, Yang},
  booktitle={Proceedings of the ACM Web Conference 2026},
  pages={4781--4792},
  year={2026}
}

@incollection{bodevraga2025theory,
 author={Leticia Bode and Emily K. Vraga},
 title={The Theory of Observed Correction},
 booktitle={Observed Correction: How We Can All Respond to Misinformation on Social Media},
 publisher={Oxford University Press}, year={2025}, pages={19--34}}

@article{chuai2026spread,
 author={Yuwei Chuai and Moritz Pilarski and Thomas Renault and David Restrepo-Amariles and Aurore Troussel-Cl{\'e}ment and Gabriele Lenzini and Nicolas Pr{\"o}llochs},
 title={Community-based fact-checking reduces the spread of misleading posts on {X} (formerly {Twitter})},
 journal={Nature Communications}, volume={17}, number={1}, pages={4070}, year={2026}}

@article{e2026comments,
 author={Qinyu E and Xinrui Wei and Xialin Zhang},
 title={How Comment Exposure Affects the Effectiveness of {COVID-19} Misinformation Corrections in Countering Misbeliefs on Social Media: The Role of Message Credibility and Issue Involvement},
 journal={Health Communication}, volume={41}, number={9}, pages={1519--1536}, year={2026}}

@article{moser2026hostile,
 author={Simone Moser and Stephanie B{\"u}hrer and Diana Kostikova and J{\"o}rg Matthes},
 title={From Hostile Feedback to Withdrawal: Emotional Exhaustion as a Barrier to Social Media Misinformation Correction},
 journal={Computers in Human Behavior}, year={2026}, pages={109179},
 note={Advance online publication, September 14, 2026}}

@article{communitywriter2026,
 author={Brad Miller and Jay Baxter and Jiansong Chao and Keith Coleman and Sophie Hilgard and Daniel Ortiz},
 title={Community-Driven {API} and {AI} Writer Design for Openly Scaling {Community Notes}},
 journal={arXiv preprint arXiv:2609.40067}, year={2026}}

@article{chuai2026downstream,
 author={Yuwei Chuai and Thomas Renault and Nicolas Pr{\"o}llochs and Gabriele Lenzini and Mohsen Mosleh},
 title={Community corrections have divergent downstream effects across corrected accounts},
 journal={arXiv preprint arXiv:2608.27526}, year={2026}}

@article{wojcik2022birdwatch,
 author={Stefan Wojcik and Sophie Hilgard and Nick Judd and Delia Mocanu and Stephen Ragain and M. B. Fallin Hunzaker and Keith Coleman and Jay Baxter},
 title={Birdwatch: Crowd Wisdom and Bridging Algorithms can Inform Understanding and Reduce the Spread of Misinformation},
 journal={arXiv preprint arXiv:2210.15723}, year={2022}}

@article{truong2025community,
 author={Bao Tran Truong and Siqi Wu and Alessandro Flammini and Filippo Menczer and Alexander J. Stewart},
 title={Community Notes are Vulnerable to Rater Bias and Manipulation},
 journal={arXiv preprint arXiv:2511.02615}, year={2025}}

@inproceedings{schafer2024hierarchical,
  title     = {Hierarchical Adversarial Correction to Mitigate Identity Term Bias in Toxicity Detection},
  author    = {Sch{\"a}fer, Johannes and Heid, Ulrich and Klinger, Roman},
  booktitle = {Proceedings of the 14th Workshop on Computational Approaches to Subjectivity, Sentiment, {\&} Social Media Analysis},
  year      = {2024},
  month     = aug,
  address   = {Bangkok, Thailand},
  publisher = {Association for Computational Linguistics},
  pages     = {35--51},
  url       = {https://aclanthology.org/2024.wassa-1.4/}
}

@article{cui2021pretraining,
  author = {Yiming Cui and Wanxiang Che and Ting Liu and Bing Qin and Ziqing Yang},
  title = {Pre-Training with Whole Word Masking for Chinese {BERT}},
  journal = {IEEE/ACM Transactions on Audio, Speech, and Language Processing},
  volume = {29},
  pages = {3504--3514},
  year = {2021}
}

@inproceedings{he2021deberta,
  author = {Pengcheng He and Xiaodong Liu and Jianfeng Gao and Weizhu Chen},
  title = {{DeBERTa}: Decoding-Enhanced {BERT} with Disentangled Attention},
  booktitle = {International Conference on Learning Representations},
  year = {2021}
}

@inproceedings{hu2022lora,
  author = {Edward J. Hu and Yelong Shen and Phillip Wallis and Zeyuan Allen-Zhu and Yuanzhi Li and Shean Wang and Lu Wang and Weizhu Chen},
  title = {{LoRA}: Low-Rank Adaptation of Large Language Models},
  booktitle = {International Conference on Learning Representations},
  year = {2022}
}

@article{yang2025qwen3,
  author = {An Yang and Anfeng Li and Baosong Yang and others},
  title = {{Qwen3} Technical Report},
  journal = {arXiv preprint arXiv:2505.09388},
  year = {2025}
}

@misc{qwen36,
  author = {{Qwen Team}},
  title = {{Qwen3.6-35B-A3B}: Agentic Coding Power, Now Open to All},
  year = {2026},
  month = apr,
  url = {https://qwen.ai/blog?id=qwen3.6-35b-a3b}
}

\appendix

\setcounter{dbltopnumber}{3}

\section{Limitations, Ethical Considerations, and Future Work}

\dataset{} provides a foundation for studying correction context understanding, while publicly retrievable Weibo records limit access to recommendation mechanisms, reviewer workflows, and policy rationales. Engagement windows vary, and display status reflects collection-time observations, limiting precise assessment of exposure and audience response. The five policy events support descriptive analysis but cannot isolate policy effects from concurrent changes in platform activity. These boundaries motivate platform collaborations to examine internal processes, longitudinal and controlled studies to assess policy effects, and cross-platform evaluations to test generalizability and extend correction context understanding.

\textbf{Ethical considerations.}
Public availability does not eliminate privacy risks, particularly when linking annotations, source posts, and reader comments. Dataset sharing and downstream use should protect contributor anonymity and minimize re-identification risks. Given the observed model biases and prediction failures, consequential decisions about publication, moderation, and platform governance should remain subject to human verification and oversight.


\section{Data and Annotation}
\label{sec:labeling-workflow}

\subsection{Samples and Splits}
Each of T1--T5 contains 20,000 training, 2,000 validation, and 2,000 test instances. T1--T3 share 1,822 test annotations with source text, human type and quality references, and observed engagement. T1 adds 178 non-annotations; T2 adds 178 quality-rated annotations; T3 adds 178 undisplayed annotations. T1/T2 retain their training/validation sets. T3 uses a random split separated by source-post and identical-input groups and labels derived from archived engagement counts. T4/T5 share random comment-level splits, with 4,000 toxic comments in training and 414 in testing.

\paragraph{Additional Data Distributions.}
\label{sec:domain-coverage}\label{sec:interaction-appendix}
Zero engagement occurs in 48.6\% of the revised 24,000-instance T3 sample and 27.7\% of its test set. The shared test cohort prioritizes displayed annotations with human quality references, so its composition does not represent all submissions. These are benchmark-sample statistics, rather than archive-wide estimates.

\subsection{Annotation}
Six experienced Weibo Annotations users independently labeled items after codebook discussion. Quality references average ratings within each dimension (Table~\ref{tab:quality-codebook}); one author resolved categorical disagreements. Stance targets the annotation, with rhetorical rejection coded as opposition. Toxicity captures verbal attacks independently of stance.

We adopted the prompt-aligned Qwen3-4B labeling procedure from prior work~\cite{weiboChiStudy}, which used alignment/test splits of 1,350/150 posts and 1,800/200 comments and reported quality QWK of 0.662 against human references. The frozen procedure labeled benchmark training/validation data with manual spot checks. Benchmark test labels were human-provided for T1/T2/T4/T5 and engagement-derived for T3. Prompt alignment applied to the labeling procedure, separate from zero-shot model evaluation.

\begin{table*}[!t]
\caption{\textbf{T2: Five-point quality codebook.} Scores 2 and 4 lie between the adjacent anchors. Adapted from the five-dimensional codebook in prior Weibo Annotation research~\cite{weiboChiStudy}.}
\label{tab:quality-codebook}
\centering\normalfont\small
\setlength{\tabcolsep}{4pt}\renewcommand{\arraystretch}{1.12}
\begin{tabularx}{\textwidth}{@{}p{1.8cm}XXX@{}}
\toprule
Dimension & 1: Low & 3: Intermediate & 5: High \\
\midrule
Relevance & Unrelated or tangential to the focal claim. & Addresses part of the claim or context. & Directly addresses the claim or pertinent context. \\
Evidence support & Bare assertion with little identifiable support. & Some concrete evidence or reasoning, with gaps. & Specific, traceable support clearly linked to the conclusion. \\
Objectivity & Unsupported inference or substantial overstatement. & Mostly factual, with some unqualified claims. & Claims fit the visible material and are appropriately qualified. \\
Helpfulness & Little useful information beyond repetition. & Useful clarification that leaves important uncertainty. & Materially helps readers assess or verify the source post. \\
Fluency & Difficult to follow because of language or structure. & Understandable despite noticeable ambiguity or awkwardness. & Clear, coherent, and easy to understand. \\
\bottomrule
\end{tabularx}
\end{table*}

\section{Experimental Setup (T1--T5)}
\label{sec:repro}\label{sec:encoder-extra}

\subsection{Supervised Models}
XGBoost and Random Forest use character TF-IDF features fitted on training data and reduced by SVD. Validation selects model settings by Macro-F1 for classification and MAE for quality regression.

The encoder checkpoints are \path{hfl/chinese-roberta-wwm-ext-large} and \path{IDEA-CCNL/Erlangshen-DeBERTa-v2-320M-Chinese}. They are fine-tuned separately for each task, with validation-based selection and early stopping.

Qwen3-4B uses separate LoRA adapters for T1, T2, and T3, and a shared adapter for T4/T5. Training applies loss to completion tokens only; validation completion loss selects checkpoints. The base and adapted models share prompts and decoding settings.

\subsection{LLMs Setting}
We access Qwen3.6, GPT-6-Astra, Gemini-3.1, and DeepSeek-V4.1 through DMXAPI\footnote{\url{https://www.dmxapi.cn/}}. T1--T5 use fixed Chinese zero-shot prompts at temperature zero. Thinking is disabled for Qwen3.6 and DeepSeek-V4.1; the other models use their default reasoning settings. Local Qwen3-4B models use non-thinking greedy decoding.

\subsection{Scoring}
T3 LLMs predict categories and $\ln(1+\mathrm{engagement})$; ML/LM baselines fit separate classifiers and regressors. Invalid classifications count as errors. T2/T3 regression uses valid, unrounded outputs, with MAE, RMSE, and Pearson $r$; T2 averages dimension-wise metrics. Results average three complete full-test runs. For T1--T3, LoRA uses one predetermined seed-17 adapter and three actual inference runs; T4/T5 retain three trained adapters. Table~\ref{tab:model-variance} summarizes observed run variability.

The analyses in Figure~\ref{fig:capability-diagnostic} use common cases across the four zero-shot LLMs, with annotation types defined by human labels. Quality averages five dimension scores, and engagement is $\ln(1+e)$, where $e$ is total engagement. In panel (c), cases are grouped into half-point bins using observed or each model's predicted quality scores separately; Pearson $r$ is computed from unbinned cases. Model statistics, including correlations, are calculated within each run and then averaged over three runs. The final heatmap row gives an equal-weight average across the four models.

\begin{figure*}[!t]
\centering\includegraphics[width=0.9\textwidth]{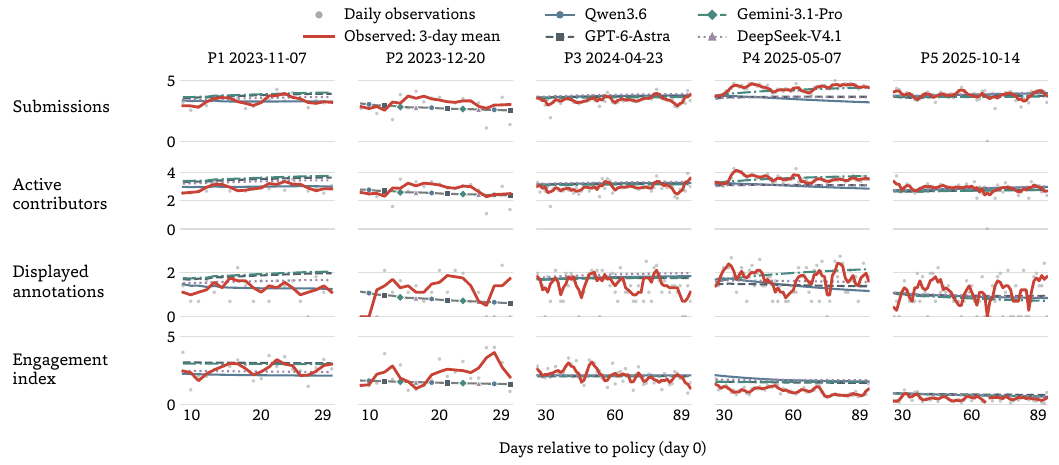}
\vspace{-1em}
\caption{\textbf{T7: All policy--outcome trajectories.} Daily observations, centered three-day means, and deterministic ABM forecasts across five policies and four outcomes. Counts use $\log(1+x)$; the engagement index retains its original scale. Scales are shared within rows, and the policy date is day 0. P2 forecast paths coincide.}
\Description{Five policy columns and four outcome rows compare daily observations with deterministic ABM forecasts. All model conditions produce the same P2 forecast paths.}
\label{fig:policy-trajectories}
\end{figure*}

\section{Policy Evaluation (T6--T7)}
\label{sec:policy-evaluation}\label{sec:abm-protocol}

\subsection{Measures and Windows}
\label{sec:simulation-implementation}
T6 and T7 track four measures: submissions, active contributors, displayed annotations, and engagement. Daily engagement averages $\log(1+e_i)$ over annotations submitted that day; display counts reflect status at collection. Table~\ref{tab:policy-windows} reports the windows and T6 reference labels. The policy date corresponds to day 1 in the table and day 0 in trajectory plots.

\begin{table}[!ht]
\caption{\textbf{Policy windows, T7 stages, and T6 reference labels.} All windows are inclusive. Baseline and history windows precede the policy. Label pairs indicate short-/long-term outcomes: I (increase), S (stable), and D (decrease).}
\label{tab:policy-windows}
\centering\small
\setlength{\tabcolsep}{3pt}
\renewcommand{\arraystretch}{1.08}
\begin{tabular*}{\columnwidth}{@{\extracolsep{\fill}}lccccc@{}}
\toprule
Stage/outcome & P1 & P2 & P3 & P4 & P5 \\
\midrule
T6 baseline & 19 days & \multicolumn{4}{c}{28 days} \\
T6 short-term & \multicolumn{5}{c}{1--14} \\
T6 long-term & \multicolumn{5}{c}{29--90} \\
\midrule
T7 history & \multicolumn{2}{c}{30 days}
           & \multicolumn{3}{c}{90 days} \\
T7 initial fit & \multicolumn{2}{c}{1--6}
               & \multicolumn{3}{c}{1--20} \\
T7 selection & \multicolumn{2}{c}{7--10}
             & \multicolumn{3}{c}{21--30} \\
T7 refit & \multicolumn{2}{c}{1--10}
         & \multicolumn{3}{c}{1--30} \\
T7 forecast & \multicolumn{2}{c}{11--30}
            & \multicolumn{3}{c}{31--90} \\
\midrule
Submissions & I/I & D/S & D/D & I/I & D/S \\
Active contributors & I/I & S/S & D/D & I/I & D/D \\
Displayed annotations & I/I & S/I & S/D & I/I & I/I \\
Engagement & I/S & S/S & S/S & S/D & S/D \\
\bottomrule
\end{tabular*}
\end{table}

\subsection{Policy Outcome QA (T6)}
\label{sec:policy-description}
\paragraph{Reference labels.}
For each policy event, Bayesian dynamic regression fitted to pre-policy data estimates counterfactual activity in the absence of the policy. We compare observed and counterfactual means over each outcome window (days 1--14 and 29--90). Relative differences within $\pm20\%$ are labeled \emph{stable}, with larger positive or negative differences labeled \emph{increase} or \emph{decrease}. When the counterfactual mean is zero, positive observed activity indicates an increase, while zero indicates stability. The threshold was fixed after inspecting class balance, yielding 11 decrease, 13 stable, and 16 increase labels.


\paragraph{Forecasting procedure.}
LLMs receive the policy text, baseline means, seven-day changes, and up to 90 days of pre-policy observations, with no post-policy data. Twelve human forecasters familiar with Weibo Annotations answer the same 40 questions using identical materials. Scores are averaged over three runs per LLM and across the 12 human forecasters. The questions share five policy events and overlapping outcome windows.

\subsection{Policy Trajectory Simulation (T7)}
\paragraph{Model and inputs.}
The shared ABM groups contributors by prior activity and display history and tracks time since their last contribution~\cite{ren2014community}. LLMs receive policy text, pre-policy history, and observations through the calibration cutoff. They propose initial and sustained changes to return, entry, submission, display, and engagement behavior~\cite{capstick2024autoelicit}.

\paragraph{Calibration and forecasting.}
We aggregate three LLM proposals to initialize parameters and priors. Calibration fits observed activity with a penalty for departures from the proposed parameters. Settings are selected by standardized MAE on the selection interval, then refitted on the full calibration prefix~\cite{muric2020socialsim,hyndman2021forecasting}. Held-out forecasts use deterministic ABM expectations with frozen parameters.

\paragraph{Evaluation.}
MAE, RMSE, Pearson $r$, and DTW compare forecasts with raw held-out observations, retaining valid zeros and omitting missing measurements. DTW uses the square root of minimum accumulated squared differences; contiguous observed blocks are aligned separately and combined by root-sum-squares. Outcome-specific metrics average equally over the five policies (Table~\ref{tab:policy-simulation}).

\section{Supplementary Results}
\label{sec:supplementary-results}

\paragraph{Quality Assessment (T2).}
Table~\ref{tab:quality-dimensions} reports errors by quality dimension, complementing the aggregate results in Table~\ref{tab:main-results}.

\begin{table}[!ht]
\caption{\textbf{Quality assessment by dimension (T2).} MAE on the revised 2,000-note test, averaged over three runs for relevance (Rel.), evidence support (Evid.), objectivity (Obj.), helpfulness (Help.), and fluency (Flu.). Formatting follows Table~\ref{tab:main-results}.}
\label{tab:quality-dimensions}
\definecolor{resultshade1}{HTML}{FCFBFF}
\definecolor{resultshade2}{HTML}{F3EFFC}
\definecolor{resultshade3}{HTML}{E7DFF9}
\definecolor{resultshade4}{HTML}{D7CAF6}
\definecolor{resultshade5}{HTML}{C6B5F2}
\centering\normalfont\small

\setlength{\tabcolsep}{3pt}\renewcommand{\arraystretch}{1.08}
\begin{tabular*}{\columnwidth}{@{\extracolsep{\fill}}lrrrrr@{}}
\toprule
Model & Rel.$\downarrow$ & Evid.$\downarrow$ & Obj.$\downarrow$ & Help.$\downarrow$ & Flu.$\downarrow$ \\
\midrule
XGBoost & \cellcolor{resultshade2}.805 & \cellcolor{resultshade2}1.270 & \cellcolor{resultshade3}.835 & \cellcolor{resultshade3}.933 & \cellcolor{resultshade2}.952 \\
Random Forest & \cellcolor{resultshade1}.832 & \cellcolor{resultshade2}1.326 & \cellcolor{resultshade2}.887 & \cellcolor{resultshade3}.955 & \cellcolor{resultshade1}.968 \\
\addlinespace[2pt]
RoBERTa & \cellcolor{resultshade3}.559 & \cellcolor{resultshade4}\underline{.955} & \cellcolor{resultshade5}\textbf{.608} & \cellcolor{resultshade4}\underline{.829} & \cellcolor{resultshade4}.788 \\
DeBERTa & \cellcolor{resultshade4}.550 & \cellcolor{resultshade5}\textbf{.951} & \cellcolor{resultshade4}\underline{.617} & \cellcolor{resultshade4}.835 & \cellcolor{resultshade3}.793 \\
\addlinespace[2pt]
Qwen3-4B & \cellcolor{resultshade1}1.752 & \cellcolor{resultshade3}1.080 & \cellcolor{resultshade1}1.536 & \cellcolor{resultshade2}.975 & \cellcolor{resultshade1}1.617 \\
Qwen3-4B LoRA & \cellcolor{resultshade3}.662 & \cellcolor{resultshade4}.981 & \cellcolor{resultshade4}.656 & \cellcolor{resultshade5}\textbf{.812} & \cellcolor{resultshade3}.799 \\
\addlinespace[2pt]
Qwen3.6 & \cellcolor{resultshade2}.747 & \cellcolor{resultshade1}1.536 & \cellcolor{resultshade1}1.451 & \cellcolor{resultshade1}1.194 & \cellcolor{resultshade1}1.001 \\
GPT-6-Astra & \cellcolor{resultshade4}\underline{.533} & \cellcolor{resultshade1}1.719 & \cellcolor{resultshade1}1.689 & \cellcolor{resultshade1}.989 & \cellcolor{resultshade5}\textbf{.558} \\
Gemini-3.1 & \cellcolor{resultshade5}\textbf{.394} & \cellcolor{resultshade1}1.337 & \cellcolor{resultshade3}.756 & \cellcolor{resultshade2}.984 & \cellcolor{resultshade4}\underline{.596} \\
DeepSeek-V4.1 & \cellcolor{resultshade1}1.032 & \cellcolor{resultshade3}1.242 & \cellcolor{resultshade2}1.369 & \cellcolor{resultshade1}1.051 & \cellcolor{resultshade2}.932 \\
\bottomrule
\end{tabular*}
\end{table}

\begin{table}[t!]
\caption{\textbf{Run variability.} Mean normalized within-task sample variance ($\times10^{-4}$) across repeated tasks.}
\label{tab:model-variance}\centering\normalfont\small\normalfont\small

\begin{tabular*}{0.8\columnwidth}{@{\extracolsep{\fill}}lr@{}}
\toprule
Model & Mean variance \\
\midrule
XGBoost & .676 \\
Random Forest & .319 \\
RoBERTa & .996 \\
DeBERTa & 1.874 \\
Qwen3-4B & .000 \\
Qwen3-4B LoRA & .838 \\
\midrule
Qwen3.6 & 21.059 \\
GPT-6-Astra & 4.141 \\
Gemini-3.1 & 15.669 \\
DeepSeek-V4.1 & 1.434 \\
\bottomrule
\end{tabular*}
\end{table}

\paragraph{Run Variability.}
For each model, we first calculate the sample variance of its scores across three runs on each task. Table~\ref{tab:model-variance} reports the average of these task-level variances. Lower values indicate more consistent performance across the recorded runs.

We use Macro-F1 for T1/T3/T4/T6 and Toxic-F1 for T5. For T2, we divide MAE by four to account for the range of the 1--5 quality scale. Rows are grouped by task coverage: T1--T5 or T1--T6. Training seeds and repeated inference measure different sources of variation, so the table is descriptive rather than a common training-stability ranking. T7 is excluded because each model produces a single deterministic forecast for each policy.

\end{document}